\documentclass[useAMS,usenatbib]{mn2e}

\usepackage{graphicx}
\usepackage{psfig}

\newcommand\apj{ApJ}
\newcommand\apjl{ApJ}
\newcommand\apjs{ApJS}
\newcommand\aap{A$\&$A}
\newcommand\mnras{MNRAS}
\newcommand\prd{Phys.~Rev.~D}
\newcommand\physrep{Phys.~Rep.}

\title[Non-Gaussian Distribution and Clustering of Hot and Cold CMB Pixels]{Non-Gaussian Distribution and Clustering of Hot and Cold Pixels in the WMAP Five-Year Sky}
\author[Graziano Rossi et al.]
{Graziano Rossi$^{1}$\thanks{Email: graziano@kias.re.kr}, Ravi K. Sheth$^{2}$, Changbom Park$^{1}$ and Carlos Hern{\'a}ndez-Monteagudo$^{3}$\\
\\ 
$^{1}$ Korea Institute for Advanced Study, Hoegiro 87, Dongdaemun-Gu, Seoul $130-722$, Korea\\
$^{2}$ Department of Physics and Astronomy, University of Pennsylvania, 209 South $33^{rd}$ Street,
       Philadelphia, PA $19104-6396$, USA\\
$^{3}$ Max Planck Institut fuer Astrophysik, Karl-Schwarzschild
       Str. 1, D-85741 Garching bei Muenchen, Germany}
\date{\today}

\pagerange{\pageref{firstpage}--\pageref{lastpage}}
\pubyear{0000}

\begin{document}
\maketitle
\label{firstpage}



\begin{abstract}
We present measurements of the clustering of hot and cold patches in the 
microwave background sky as measured from the 
\textit{W}ilkinson \textit{M}icrowave \textit{A}nisotropy \textit{P}robe 
(WMAP) five-year data. These measurements are compared with theoretical predictions which assume 
that the cosmological signal obeys Gaussian statistics. We find
significant differences from the simplest Gaussian-based prediction.  
However, the measurements are sensitive to the fact that the noise is 
spatially inhomogeneous (e.g., because different parts of the sky 
were observed for different lengths of time). We show how to account 
for this spatial inhomogeneity when making predictions. Differences 
from the Gaussian-based expectation remain even after this more careful 
accounting of the noise. In particular, we note that
hot and cold pixels cluster differently within the same temperature thresholds at few-degree scales. 
While these findings may indicate primordial non-Gaussianity, we discuss other plausible explanations for these discrepancies.  
In addition, we find some deviations from Gaussianity 
at sub-degree scales, especially in the W band, whose origin may be associated
with extragalactic dust emission.     
\end{abstract}






\section{Introduction}

Non-Gaussianity of the Cosmic Microwave Background (CMB) is progressively
becoming a crucial probe of the inflationary dynamics,
since models beyond single field slow-roll inflation predict large distinct
non-Gaussianities above the current limit of measurement 
(Lyth et al. 2003, 2006; Bartolo et al. 2004; Dvali et al. 2004; Chen
2005; Seery \& Lidsey 2005; 
Chen \& Szapudi 2006; Seery \& Hidalgo 2006; Ling \& Wu 2008; Senatore
et al. 2008, 2009; Chen et al. 2009).
Moreover, up-to-date a number of plausible alternative early universe
models also predict skewed primordial fluctuations, and are in principle distinguishable
from other scenarios through the shape dependence of high-order
correlation functions
(Alishahiha et al. 2004; Arkani-Hamed et al. 2004; 
Creminelli \& Senatore 2007; Brandenberger 2008; Buchbinder, Khoury \& Ovrut
2008; Lehners \& Steinhardt 2008; McAllister \& Silverstein 2008).  

An even small degree of primordial non-Gaussianity would
indicate a quite different structure formation scenario from the
concordance cosmological model in which density
perturbations are assumed to be a Gaussian random field, 
and alter significantly the statistics of voids (Kamionkowski et al. 2008; Song \& Lee 2008, for example).
Imprints of primordial non-Gaussianity can manifest in the
statistical properties of the Lyman-$\alpha$ forest QSO spectra at
intermediate redshift (Viel et al. 2008), in the large-scale
distribution of neutral hydrogen (Pillepich et al. 2007),
in the reionization history
of the universe (Crociani et al. 2008), and in the abundance,
clustering and biasing of dark matter halos (Kang
et al. 2007; Carbone et al. 2008; Dalal et al. 2008; Desjacques et al. 2008;
Grossi et al. 2008; LoVerde et al. 2008; Matarrese \& Verde 2008; McDonald \& Seljak 2008; Pillepich et al. 2008;
Seljak 2008; Slosar et al. 2008; Jeong \& Komatsu 2009). 
Topology of the large scale structure is also very sensitive to the initial
conditions of the matter density field, and potentially represents a
strong and independent test of deviations from the Gaussian 
hypothesis (Park et al. 1998, 2001, 2005; Gott et al. 2008a,b; Hikage et al. 2008b).  

Hence, there is significant interest in quantifying the Gaussianity 
of the cosmic background radiation, and a multitude of non-Gaussian 
estimators have been applied so far.
In particular, data from the \textit{W}ilkinson \textit{M}icrowave \textit{A}nisotropy 
\textit{P}robe (WMAP) mission has been used to constrain primordial 
non-Gaussianity. Recent measurements of the one- and three-point 
probability functions, the bispectrum and trispectrum, the genus 
statistic and the other Minkowski functionals 
(Cabella et al. 2005; Hinshaw et al. 2007, 2008; Creminelli et al. 2007;
Gott et al. 2007;  Spergel et al. 2007; Curto et
al. 2008; Hikage et al. 2008a; Komatsu
et al. 2008; Vielva \& Sanz 2008; Pietrobon et al. 2009a) from the first three or five years of WMAP data are all 
consistent with Gaussianity.
However, a considerable number of anomalies like asymmetries,
intrinsic alignments in the data or the presence of unusual cold spots 
have also been reported
(Cruz et al. 2006; Tojeiro et al. 2006; Chiang et al. 2007; Naselsky
et al. 2007; R{\"a}th et al. 2007, 2009; Vielva et
al. 2007; Hansen et al. 2008; Dickinson et al. 2009; Diego et
al. 2009; Li et al. 2009; Pietrobon et al. 2009b; Rossmanith et al. 2009), as well as some
claims of non-Gaussianity (Jeong \& Smoot 2007; Yadav \& Wandelt 2008), 
which, if confirmed, would rule out a large class of inflationary models. 

In this paper we investigate the Gaussian hypothesis using the 
clustering statistics of pixels that lie above or below a threshold.  
For a Gaussian field, the scale dependence of this statistic, and 
its dependence on threshold, have been predicted (Jensen \& Szalay 1986).  
It turns out that a careful accounting for the effects of noise is 
crucial to performing this test. Kashlinsky et al. (2001) and Hern{\'a}ndez-Monteagudo et al. (2004) 
presented an analysis of the problem when the noise is spatially 
homogeneous. In the case of WMAP, the noise is spatially 
inhomogeneous; the present work shows how to incorporate this 
complexity into the analysis.  

The layout of the paper is as follows.
In Section \ref{noise_model} we outline 
the theoretical framework for the two-point statistics above a temperature
threshold, and show how to incorporate the complexity of inhomogeneous
noise into the two-point formalism. Some details on errorbar
calculations are left in Appendix A. 
In Section \ref{WMAP5 analysis} we test our theory against WMAP
five-year data.
We study and quantify the inhomogeneous noise
properties and show that the abundance and clustering of pixels 
(both scale and dependence on threshold temperature) appears to be 
inconsistent with the Gaussian prediction if one ignores the fact 
that the noise is spatially inhomogeneous. We compare the number
density of pixels above threshold and the one- and two-point
statistics measurements with the theory, allowing for inhomogeneous noise.   
While this improves the agreement between 
measurements and predictions, discrepancies remain.  
Although this may be an indication of primordial non-Gaussianity, 
we discuss a variety of other possibilities, among which the effect
of smoothing the map. 
We inspect all the WMAP five-year channels, but we present in
the main text results only for the W1 differencing assembly (DA),
and leave in Appendix B results for other DA's. 
A final section summarizes our findings, and highlights ongoing and future studies.    



\section{Theoretical Model} \label{noise_model}


\subsection{Basic Notation}

Denote the observed value in a pixel by $D = T - \langle T \rangle
\equiv \delta T$, which is the sum of the true signal $s$ plus noise
$n$, both of which have mean zero.  
We consider a model in which the signal is homogeneous and may have 
spatial correlations 
whereas the noise, which is independent of 
the signal, may be inhomogeneous and have spatial 
correlations. 
By this we mean that the
rms value of the noise $\sigma_{\rm n}$ 
may fluctuate from pixel to pixel and these fluctuations may 
be correlated, and that the actual value of the noise in one pixel 
could depend of that in another.
Let $p(D)$ denote the observed one-point distribution of $D$, 
$G(s)$ the distribution of $s$, 
$p(\sigma_{\rm n})$ the distribution of the rms value of the noise 
in a pixel, 
and $g(n|\sigma_{\rm n})$ the distribution of the noise when the rms 
value of the noise is $\sigma_{\rm n}$. Note that $g(n) = \int
g(n|\sigma_{\rm n}) p(\sigma_{\rm n}) {\rm d}\sigma_{\rm n}$.
Our convention is to use capital letters for average quantities, and
lower case letters for actual (varying) quantities. Later on in the
paper we will also drop
the (understood) subscript $n$ of the rms of the noise, for clarity of notation.


\subsection{One-Point Observed Distribution}

Since the distribution of the noise $g(n)$ is independent 
of the signal $s$, 
\begin{eqnarray}
p(D) &=& \int {\rm d}s~G(s) \int {\rm d}n~g(n) \delta_{\rm D} (s+n=D) \nonumber \\
     &=& \int {\rm d}s~G(s) \int {\rm d}n \int {\rm d}\sigma_{\rm n}~g(n|\sigma_{\rm
     n})~p(\sigma_{\rm n}) \nonumber \\
     & & \times~\delta_{\rm D}(s+n=D) \nonumber \\
     &=& \int {\rm d}\sigma_{\rm n}~p(\sigma_{\rm n}) \int
     {\rm d}s~G(s)~g(D-s|\sigma_{\rm n}) \nonumber \\ 
     &=& \int  {\rm d}\sigma_{\rm n}~p(\sigma_{\rm n})~p(D|\sigma_{\rm n})
\label{pD}
\end{eqnarray}
\noindent where $\delta_{\rm D}$ is the Dirac delta and
\begin{equation} 
p(D|\sigma_{\rm n}) =\int {\rm d}s~G(s)~g(D-s|\sigma_{\rm n}). 
\label{pD_given_sigma}
\end{equation}
The variance of $D$ is:
\begin{eqnarray}
\sigma_{\rm D}^2 &\equiv& \langle D^2\rangle = \int {\rm d}D\, p(D)\, D^2 \nonumber  \\
                &=& \int {\rm d}\sigma_{\rm n}\, p(\sigma_{\rm n})
                \int {\rm d}s\,
                G(s) \, \int {\rm d}n\,g(n|\sigma_{\rm n})\, (s+n)^2 \nonumber \\
                &=& \sigma_{\rm S}^2 + \int {\rm d}\sigma_{\rm n}\, p(\sigma_{\rm
                n})\,\langle n^2|\sigma_{\rm n} \rangle \nonumber \\
                &=& \sigma_{\rm S}^2 + \sigma_{\rm N}^2 
\label{sigmaD}
\end{eqnarray}
\noindent where we have used $\sigma_{\rm S}$ to denote the rms value of the
signal, and $\sigma_{\rm N}^2$ to denote the variance of the noise upon
averaging over all pixels.
It is also straightforward to show that $\langle D \rangle = 0$,  as
expected. 

The fraction of pixels above some threshold $D_{\rm t}$ is
\begin{eqnarray}
 f(D_{\rm t}) &=& \int_{D_{\rm t}}^\infty {\rm d}D\, p(D)
        = \int {\rm d}\sigma_{\rm n}\, p(\sigma_{\rm n})\,
            \int_{D_{\rm t}}^\infty {\rm d}D\, p(D|\sigma_{\rm n}) \nonumber \\
        &=& \int {\rm d}\sigma_{\rm n}\, p(\sigma_{\rm n})\, f(D_{\rm
        t}|\sigma_{\rm n}).
\label{fDt}
\end{eqnarray}
If $\langle D|D_{\rm t}\rangle$ denotes the mean value of $D$ in such
pixels, then
\begin{eqnarray}
f(D_{\rm t})\,\langle D|D_{\rm t} \rangle &=& \int_{D_{\rm t}}^\infty
                              {\rm d}D\, p(D)\, D \nonumber \\
                              &=& \int {\rm d}\sigma_{\rm n}\, p(\sigma_{\rm
                              n})\,\int_{D_{\rm t}}^\infty {\rm d}D\,
                              p(D|\sigma_{\rm n})\, D \nonumber \\
                              &=& \int {\rm d}\sigma_{\rm n}\, p(\sigma_{\rm
                              n})\,f(D_{\rm t}|\sigma_{\rm n}) \,
                              \langle D|D_{\rm t},\sigma_{\rm n}\rangle .
\label{fDtDDt}
\end{eqnarray}


\subsection{Two-Point Observed Distribution}

Two point statistics may be computed similarly.
For two pixels separated by the angular distance $\theta$,
{\setlength\arraycolsep{0.5pt}
\begin{eqnarray}
\lefteqn{p(D_1,D_2|\theta) = \int {\rm d}s_1 \int
  {\rm d}s_2~G(s_1,s_2|\theta)
  \int {\rm d}n_1 \int {\rm d}n_2 {}} \nonumber \\ 
& & {} ~~~ \times g(n_1,n_2)~\delta_{\rm D} (s_1
  +n_1=D_1)~\delta_{\rm D} (s_2+n_2=D_2) \nonumber \\
& & {} = \int {\rm d}s_1 \int {\rm d}s_2~G(s_1,s_2|\theta) \int {\rm d}n_1
  \int {\rm d}n_2 \int {\rm d}\sigma_1 \nonumber \\
& & {} ~~~ \times \int {\rm d}\sigma_2
  ~g(n_1,n_2|\sigma_1,\sigma_2)~p(\sigma_1,\sigma_2|\theta)~\delta_{\rm D} (s_1+n_1=D_1) \nonumber \\
& & {} ~~~ \times~\delta_{\rm D} (s_2+n_2=D_2) \nonumber \\
& & {}  = \int {\rm d}\sigma_1 \int {\rm d}\sigma_2\,
  p(\sigma_1,\sigma_2|\theta)
           \int {\rm d}s_1 \int {\rm d}s_2\, G(s_1,s_2|\theta)  \nonumber  \\
& & {} ~~~ \times g(D_1-s_1|\sigma_1)\,g(D_2-s_2|\sigma_2) \nonumber  \\
& & {} = \int {\rm d}\sigma_1 \int {\rm d}\sigma_2\,
                         p(\sigma_1,\sigma_2|\theta)\,
                         p(D_1,D_2|\sigma_1,\sigma_2,\theta)
\label{pD1D2}
\end{eqnarray}}
\noindent where 
\begin{eqnarray}
 p(D_1,D_2|\sigma_1,\sigma_2,\theta) &=& \int {\rm d}s_1 \int {\rm d}s_2\,
 G(s_1,s_2|\theta) \nonumber \\
 &\times& g(D_1-s_1|\sigma_1)\,g(D_2-s_2|\sigma_2). 
\label{pD1D2|sig1sig2}
\end{eqnarray}
Notice that if we integrate over all pixels then
\begin{eqnarray}
\langle D_1D_2|\theta\rangle
   &=& \int {\rm d}D_1 \int {\rm d}D_2\,  D_1D_2\, p(D_1,D_2|\theta) \nonumber \\
   &=&  \int {\rm d}s_1 \int {\rm d}s_2\, G(s_1,s_2|\theta)\,s_1s_2 \nonumber \\
   && + \int {\rm d}n_1 \int {\rm d}n_2~g(n_1,n_2| \theta)~n_1 n_2  \nonumber \\
   &=& \langle s_1s_2|\theta \rangle + \langle n_1n_2|\theta \rangle
   \nonumber \\
   &\equiv& C_{\rm S}(\theta) + C_{\rm N}(\theta) = C(\theta),
\label{D1D2}
\end{eqnarray}
\noindent where
\begin{equation}
\langle s_1s_2|\theta \rangle = \int {\rm d}s_1 \int {\rm d}s_2\,
G(s_1,s_2|\theta)\,s_1s_2 \equiv C_{\rm S}(\theta)
\end{equation}
\noindent and
\begin{eqnarray}
\langle n_1n_2|\theta \rangle
   &=& \int {\rm d}n_1 \int {\rm d}n_2~g(n_1,n_2| \theta)~n_1 n_2 \nonumber\\
&=& \int {\rm d}n_1 \int {\rm d}n_2
   \int_{0}^{\infty} {\rm d}\sigma_1 \int_{0}^{\infty} {\rm d}\sigma_2
   ~p(\sigma_1,\sigma_2|\theta) \nonumber \\
&&\times~g(n_1,n_2|\sigma_1,\sigma_2,\theta)~n_1
   n_2 \nonumber\\
&=& \int_{0}^{\infty} {\rm d}\sigma_1 \int_{0}^{\infty} {\rm d}\sigma_2~p(\sigma_1,\sigma_2|\theta)
   \int {\rm d}n_1 \int {\rm d}n_2\nonumber \\
&&\times~g(n_1,n_2|\sigma_1,\sigma_2,\theta)~n_1
   n_2 \nonumber\\
&=& \int_{0}^{\infty} {\rm d}\sigma_1 \int_{0}^{\infty} {\rm d}\sigma_2
   ~p(\sigma_1,\sigma_2|\theta)~\langle n_1
   n_2|\sigma_1,\sigma_2,\theta \rangle \nonumber\\
&\equiv& C_{\rm N}(\theta). 
\label{n1n2_cov}
\end{eqnarray}
\noindent This is because the signal and noise are uncorrelated, and so
$\langle s n|\theta\rangle = 0$ for all $\theta>0$. Clearly, if the
noise is not spatially correlated, then $\langle n_1n_2|\theta
\rangle = 0$ and $\langle D_1D_2|\theta\rangle \equiv C_{\rm S}(\theta)$.

If the rms values of the noise are perfectly correlated,
meaning $\sigma_1=\sigma_2$, then it is convenient to write
$p(\sigma_1,\sigma_2|\theta) = p(\sigma_1)\, p(\sigma_2|\sigma_1, \theta)$
and replace $p(\sigma_2|\sigma_1, \theta)$ with a delta function centered
on $\sigma_1=\sigma_{\rm n}$.
One then averages over the distribution of $\sigma_{\rm n}$, namely:
\begin{eqnarray}
p(D_1,D_2|\theta) &=& \int {\rm d}\sigma_1 \int {\rm d}\sigma_2\,
                         p(\sigma_1)
                         ~\delta_{\rm D}(\sigma_1 = \sigma_2 \equiv
                         \sigma_{\rm n}) \nonumber \\
                         && \times~p(D_1,D_2|\sigma_1,\sigma_2,\theta) \nonumber \\
                  &=&\int {\rm d}\sigma_{\rm n}~p(\sigma_{\rm n})
                    ~p(D_1,D_2|\sigma_{\rm n},\sigma_{\rm n}, \theta).
\label{pD1D2_bis}
\end{eqnarray}
\noindent If, in addition, one replaces $p(\sigma_{\rm n})$ with a delta function
centered on $\sigma_{\rm N}$, then $p(D_1,D_2|\theta) \equiv
p(D_1,D_2|\sigma_{\rm N},\sigma_{\rm N}, \theta)$, and the homogeneous case is recovered.
The noise distributions $p(\sigma_{\rm n})$ and $p(\sigma_1,\sigma_2|\theta)$ are
measured directly from the WMAP five-year data.


\subsection{Two-Point Function above Threshold and Inhomogeneous Noise} \label{two_point_statistics} 

The two-point unweighted correlation function $\xi(\theta)$ estimates the excess of probability of finding a pair
at a distance $\theta$, compared to a random catalog (e.g. Jensen \& Szalay, 1986).
Similarly, one can define the two-point weighted correlation function $W(\theta)$, 
which differs from $\xi(\theta)$ only because each member of the pair is now weighted by some (average) mark.  
In this study, the mark is given by $\delta T/ \langle \delta T \rangle \equiv D/ \langle D \rangle$. 
Two-point correlation functions of regions above some threshold are
readily obtained by integration above the threshold level. 

Within the formalism previously outlined, the two
point correlation function of pixels above some threshold $D_{\rm t}$ is
\begin{equation}
 1 + \xi_{\rm t}(\theta) = \int_{D_{\rm t}}^\infty {\rm d}D_1
                     \int_{D_{\rm t}}^\infty {\rm d}D_2\,
                     {p(D_1,D_2|\theta)\over f(D_{\rm t})^2}
\label{xitheta}
\end{equation}
and the associated weighted function is
\begin{equation}
 1 + W_{\rm t}(\theta) = \int_{D_{\rm t}}^\infty {\rm d}D_1
                     \int_{D_{\rm t}}^\infty {\rm d}D_2\,
                     {D_1 D_2\over \langle D|D_{\rm t}\rangle^2}\,
                     {p(D_1,D_2|\theta)\over f(D_{\rm t})^2},
\label{Wtheta}
\end{equation}
\noindent where $p(D_1,D_2|\theta)$, $\langle D|D_{\rm t} \rangle$ and
$f(D_{\rm t})$ have been previously defined.

To gain intuition, it is helpful to define
\begin{equation}
 1 + \xi_{\rm t}(\theta|\sigma_1,\sigma_2)
  = \int_{D_{\rm t}}^\infty {\rm d}D_1 \int_{D_{\rm t}}^\infty {\rm d}D_2\,
        {p(D_1,D_2|\sigma_1,\sigma_2,\theta)\over
         f(D_{\rm t}|\sigma_1)f(D_{\rm t}|\sigma_2)}
\end{equation}
and 
\begin{eqnarray}
 1 + W_{\rm t}(\theta|\sigma_1,\sigma_2)&=&\int_{D_{\rm t}}^\infty
                     {\rm d}D_1 \int_{D_{\rm t}}^\infty {\rm d}D_2\,
                     {D_1 D_2\over \langle D|D_{\rm t},\sigma_1\rangle\,
                     \langle D|D_{\rm t},\sigma_2\rangle} \nonumber \\
         &\times& {p(D_1,D_2|\sigma_1,\sigma_2,\theta)\over
                      f(D_{\rm t}|\sigma_1)\,f(D_{\rm t}|\sigma_2)},
\end{eqnarray}
in terms of which
\begin{eqnarray}
 1 + \xi_{\rm t}(\theta) &=& \int {\rm d}\sigma_1 \int {\rm d}\sigma_2\,
                       p(\sigma_1,\sigma_2|\theta)\,
                       {f(D_{\rm t}|\sigma_1)\,f(D_{\rm t}|\sigma_2)\over
                       f(D_{\rm t})^2} \nonumber \\
   &\times& \Bigl[1 + \xi_{\rm t}(\theta|\sigma_1,\sigma_2)\Bigr]
\label{xiintuit}
\end{eqnarray}
and
\begin{eqnarray}
1 + W_{\rm t}(\theta) &=& \int {\rm d}\sigma_1 \int {\rm d}\sigma_2\,
                       p(\sigma_1,\sigma_2|\theta) {f(D_{\rm
                       t}|\sigma_1)\langle D|D_{\rm t},\sigma_1\rangle
                        \over f(D_{\rm t})\, \langle D|D_{\rm t}\rangle} \nonumber \\
&\times& {f(D_{\rm t}|\sigma_2)\langle D|D_{\rm t},\sigma_2\rangle
                        \over f(D_{\rm t}) \langle D|D_{\rm t}\rangle} \,
                     \Bigl[1 + W_{\rm t}(\theta|\sigma_1,\sigma_2)\Bigr].
\label{Wiintuit}
\end{eqnarray}
\noindent This shows that $\xi_{\rm t}$ and $W_{\rm t}$ can be thought of as weighted
averages over the values at fixed $\sigma_1$ and $\sigma_2$.
This re-writing shows clearly that the key quantity of interest
are $p(D_1,D_2|\sigma_1,\sigma_2,\theta)$, which is the convolution
of the signal and noise distributions, and $p(D|\sigma_{\rm n})$.
This rewriting also allows one to study two limiting cases.
If the rms values of the noise are perfectly correlated,
meaning $\sigma_1=\sigma_2\equiv \sigma_{\rm n}$, then:
\begin{eqnarray}
 1 + \xi_{\rm t}(\theta) &=& \int {\rm d}\sigma_{\rm
                   n}\,p(\sigma_{\rm n})
                   \int_{D_{\rm t}}^\infty {\rm d}D_1 \int_{D_{\rm
                   t}}^\infty {\rm d}D_2
                   \nonumber \\
        &\times& {p(D_1,D_2|\sigma_{\rm n},\sigma_{\rm n},\theta)\over
                   f(D_{\rm t})^2}
\end{eqnarray}
and 
\begin{eqnarray}
 1 + W_{\rm t}(\theta) &=& \int {\rm d}\sigma_{\rm
                   n}\,p(\sigma_{\rm n})
                   \int_{D_{\rm t}}^\infty {\rm d}D_1 \int_{D_{\rm
                   t}}^\infty {\rm d}D_2\,
        {D_1 D_2 \over \langle D|D_{\rm t} \rangle^2} \nonumber \\
	&\times& {p(D_1,D_2|\sigma_{\rm n},\sigma_{\rm n},\theta)\over
                   f(D_{\rm t})^2}.
\end{eqnarray}
If, in addition, one replaces $p(\sigma_{\rm n})$ with a delta function
centered on $\sigma_{\rm N}$, then
\begin{equation}
 1 + \xi_{\rm t}(\theta) =
                   \int_{D_{\rm t}}^\infty {\rm d}D_1 \int_{D_{\rm
                   t}}^\infty {\rm d}D_2\,
        {p(D_1,D_2|\sigma_{\rm N},\sigma_{\rm N},\theta)\over f(D_{\rm
                   t})^2}
\end{equation}
and 
\begin{eqnarray}
 1 + W_{\rm t}(\theta) &=& \int_{D_{\rm t}}^\infty {\rm d}D_1
        \int_{D_{\rm t}}^\infty {\rm d}D_2\,
        {D_1 D_2 \over \langle D|D_{\rm t} \rangle^2} \noindent
        \nonumber \\
	&\times& {p(D_1,D_2|\sigma_{\rm N},\sigma_{\rm N},\theta)\over
        f(D_{\rm t})^2}.
\end{eqnarray}

We are particularly interested in the case where the signal
$G(s_1,s_2|\theta)$ is bivariate Gaussian with
$\langle s_1^2\rangle = \langle s_2^2\rangle = \sigma_{\rm S}^2$,
and $\langle s_1s_2|\theta\rangle = C_{\rm S}(\theta)$
and the noise $g(n|\sigma_{\rm n})$ is Gaussian with rms $\sigma_{\rm n}$.
Then
{\setlength\arraycolsep{0pt}
\begin{eqnarray}
\lefteqn{p(D_1,D_2|\sigma_1,\sigma_2,\theta) = \frac{1}{2 \pi \sqrt{||C||} }
 e^{-\frac{1}{2}D^{\rm T} \cdot C^{-1} \cdot D} {}} \\
& & {} =  {1\over 2 \pi \sigma_{\rm D}^2 \sqrt{\alpha_1 \alpha_2 - w_\theta^2} }\
    {\rm exp}\left\{-{\alpha_2 D_1^2 + \alpha_1 D_2^2 - 2w_\theta\, D_1D_2\over
               2\sigma_{\rm D}^2\,(\alpha_1 \alpha_2 -
 w_\theta^2)}\right\} \nonumber
\label{pD1D2_gauss}
\end{eqnarray}}
\noindent with $\alpha_1 = (\sigma_{\rm S}^2 + \sigma_1^2)/\sigma_{\rm D}^2$,
$\alpha_2  = (\sigma_{\rm S}^2 + \sigma_2^2)/\sigma_{\rm D}^2$, and 
\begin{equation}
w_\theta = {C_{\rm S}(\theta) + 
C_{\rm N}(\theta) \over \sigma_{\rm D}^2},
\end{equation}
\begin{equation}
C_{\rm S}(\theta) = \sum_{\rm \ell} {(2\ell+1)\over 4\pi}\,
             C_{\rm \ell}\, W^{\rm WMAP}_ {\rm \ell}\,
                      W^{\rm smooth}_{\rm \ell}\
             P_{\rm \ell}^0(\cos \theta),
\label{ctheta_eq}
\end{equation}
\begin{equation}
C_{\rm N}(\theta) = \sum_{\rm \ell} {(2\ell+1)\over 4\pi}\,
             C_{\rm \ell}^{\rm N}\,
                      W^{\rm smooth}_{\rm \ell}\
             P_{\rm \ell}^0(\cos \theta),
\label{ctheta_noise}
\end{equation}
\noindent where $C$ is the covariance matrix of the temperature field, 
$C_{\rm \ell}^{\rm N}$ the power spectrum of the noise map, 
$W^{\rm WMAP}$ the \textit{WMAP} window function and $W^{\rm smooth}$ the
additional smoothing due to finite pixel size, 
mask influence and an optional additional Gaussian beam smoothing.

If the noise is spatially uncorrelated, then clearly $C_{\rm
  N}(\theta) = 0$ and therefore
$w_\theta \equiv C_{\rm S}(\theta)/\sigma_{\rm D}^2$.

In the approximation where $\sigma_1=\sigma_2$, rms noise varies
spatially on scales much larger than those of interest, then
$\alpha_1=\alpha_2$.  The ``standard'' approximation, rms noise
independent of position, has $\alpha_1=\alpha_2=1$ and we recover the
Hern{\'a}ndez-Monteagudo et al. (2004) formula.

Uncertainties in the
correlation functions above threshold are estimated from
the optimal variance limit, containing
cosmic variance, instrumental noise (Knox 1995; 
Hern{\'a}ndez-Monteagudo et al. 2004), and finite
binsize effects. See Appendix A for more details. 



\section{Analysis of WMAP 5-Year Data} \label{WMAP5 analysis}


\subsection{Dataset and Pipeline}

The data from the five years of the WMAP mission are 
available online at \textit{L}egacy \textit{A}rchive for \textit{M}icrowave 
\textit{B}ackground \textit{DA}ta (LAMBDA) website 
(http://lambda.gsfc.nasa.gov/), NASA's CMB Thematic Data Center. 
For the purposes of this study, we use the WMAP five-year full 
resolution ``foreground-reduced'' coadded sky maps (produced by
performing a weighted pixel-by-pixel mean of the five single year
maps, see Hinshaw et al. 2008) for the W (the 94 GHz channel), V (the
61 GHz channel) and Q (the 41 GHz channel)  
differencing assemblies. 
We show results for the W1 channel in the main text,
and leave in Appendix B major results for all the other individual DA's. 
The data consist of four fields for each pixel: 
\begin{enumerate}
 \item the thermodynamic temperature in mK, 
 \item the Q Polarization temperature in mK, 
 \item the U Polarization temperature in mK, and 
 \item the effective number of observations, $N_{\rm obs}$.
\end{enumerate}
The maps are provided in the HEALPix scheme (G\'orski et al. 1999) at 
a resolution of $N_{\rm side}=512$, giving a total of 3145728 pixels 
separated on average by $\theta_{\rm pix}=6.87'$.  
The maps have been cleaned of galactic foreground emission using external
templates, as explained in Gold et al. (2008).
It is indeed necessary to mask out regions of strong foreground emission.
Of the different masks provided from the LAMBDA Legacy Archive, 
which allow for the selective exclusion of portions of 
the sky at different flux levels, we chose the most conservative, 
KQ75 -- essentially an extension and improvement of the
standard KP0 cut in the 3-yr data release. 
This choice removes about 28\% of the pixels.  
Point sources are masked based on a combination of external catalog
data and WMAP-detected sources (Wright et al. 2008). 
Note that point sources are the largest astrophysical contaminant to
the temperature power spectrum.

Our theory predictions (Section \ref{noise_model}) depend on the TT power spectrum, 
on the shape of the WMAP beam, and on the noise properties.
We used the beam and window transfer functions from the five-year WMAP 
data, which have been significantly improved with the 5yr data
release; models of the instrument gain and beam response are now 
accurate to better than $\sim 0.6\%$, and errors in the five-year beam
transfer functions are reduced by a factor of $\sim 2$ as compared to
the three-year analysis (Hill et al. 2008).
The WMAP5 TT power spectrum we use comes from a weighted combination 
of 153 individual cross-power spectra, and has been improved by using
a Gibbs-based maximum likelihood estimate for $l \le 32$ (Dunkley et
al. 2008) and a pseudo $C_{\rm \ell}$ estimate for higher $l$ (Nolta et al. 2008).  
The pixel-pixel covariance matrix is essentially diagonal 
(Hinshaw et al. 2003, 2008),
so the rms noise in pixel $p$ is 
$\sigma(p)=\sigma_0/\sqrt{N_{\rm obs}(p)}$, where the noise per observation,
$\sigma_0$, is provided for each coadded channel.  
This value for $\sigma_0$ is obtained by averaging the 
values of the five possible year-by-year difference combinations 
(Jarosik et al. 2007; Hinshaw et al. 2008).  

The data reduction process should be free of contamination induced by
foregrounds, and should be insensitive to monopole and dipole moments.
At $N_{\rm side}=512$, the generic data pipeline involves the
following main steps:
\begin{enumerate}
\item dipole and monopole removal outside KQ75;
\item selection of pixels above a temperature threshold;
\item HEALPix coordinate and pixel-noise assignment.
\end{enumerate}


\subsection{Inhomogeneous Noise Properties}

\begin{figure}
\begin{center}
\includegraphics[angle=0,width=0.45\textwidth]{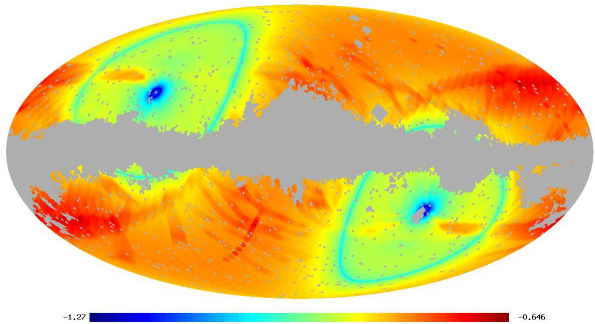}
\caption{Noise per pixel distribution at
         $N_{\rm side}=512$ for the five-year coadded WMAP 
         data (W1 channel), after application of the KQ75 mask, in
         Mollweide projection. The scale is logarithmic, in mK units
         -- i.e. $\log_{\rm 10}(\sigma_{\rm n})$ is shown.}
\label{psigma_map}
\end{center}
\end{figure}

\begin{figure}
\begin{center}
\includegraphics[angle=0,width=0.45\textwidth]{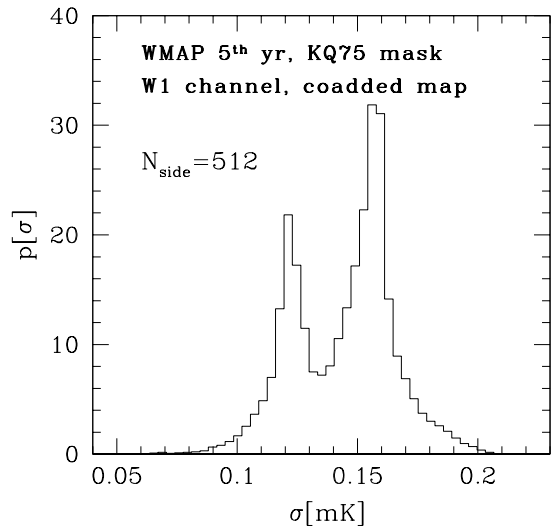}
\caption{Distribution of the rms noise-per-pixel at $N_{\rm side}=512$ for the WMAP5 
 W1 coadded channel, after application of the KQ75 mask.}
\label{psigma_hist}
\end{center}
\end{figure}

The distribution of the rms noise values in a pixel at $N_{\rm side}=512$ 
is shown in Mollweide projection in Figure \ref{psigma_map}, and as 
histogram in Figure \ref{psigma_hist}, for the WMAP5 W1 channel.  
The distribution is reasonably broad, indicating that the noise 
is inhomogeneous. This is a direct consequence of the fact that 
not all pixels were observed the same number of times. 
The evident symmetrical pattern present in Figure \ref{psigma_map} is due to
the scanning strategy of the WMAP satellite.  

\begin{figure}
 \begin{center}
 \includegraphics[angle=0,width=0.49\textwidth]{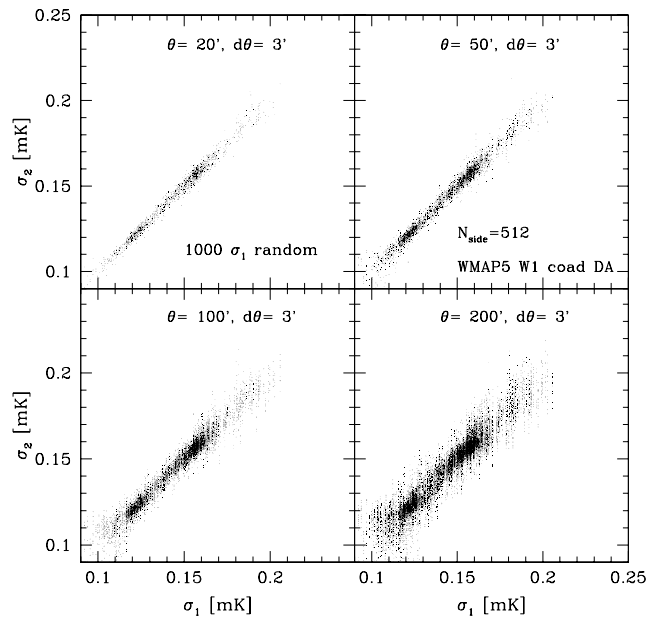}
 \caption{Joint distribution $p(\sigma_1,\sigma_2|\theta)$ at 
          $N_{\rm side}=512$ resolution for four different choices of
          $\theta$ as indicated in the panels, from the WMAP5 W1 coadded channel.
 \label{psigma12}}
 \end{center}
\end{figure}

\begin{figure}
 \begin{center}
 \psfig{figure=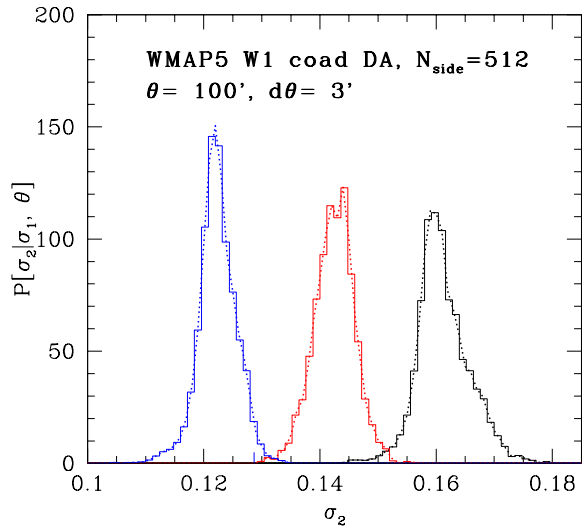,height=3.0in}
 \caption{Examples of conditional probability 
  distributions p($\sigma_2|\sigma_1, \theta$), when $\theta=100'$,
  for the WMAP5 W1 coadded channel. The panel shows results for $N_{\rm side}=512$.
  Different sets of curves (spline fits) show the 
  distribution of $\sigma_2$ for $\sigma_1$=0.122, 0.142, 0.162 mK, respectively.  
 \label{sigma12fits}}
\end{center}
\end{figure}

In particular, equations~(\ref{xiintuit}) and (\ref{Wiintuit})
show that the clustering 
of pixels separated by $\theta$ depends on the joint distribution 
of $p(\sigma_1,\sigma_2)$ given $\theta$.   
Figure~\ref{psigma12} shows an example of how the joint distribution 
$p(\sigma_1,\sigma_2|\theta)$ varies as a function of separation 
$\theta$ for the resolution $N_{\rm side}=512$.  
The four panels were constructed by randomly selecting $1000$ 
pixels, and for each selected value of $\sigma_{1}$, looking at the 
corresponding $\sigma_2$ distributions at four different angular 
distances. Note that the scatter around the one-to-one line 
increases with increasing $\theta$. This is not surprising; 
nearby pixels have similar $N_{\rm obs}$ whereas more widely 
separated ones do not (Figure \ref{psigma_map}).
Figure~\ref{sigma12fits} shows some examples of the corresponding conditional distributions 
$p(\sigma_2|\sigma_1,\theta)$.  
When inserted into equations~(\ref{xiintuit}) and~(\ref{Wiintuit}), such fits allow a 
prediction of how the clustering of pixels above a certain threshold 
temperature should depend on threshold if the signal is Gaussian and 
the noise is inhomogeneous.

\subsection{Distribution of Temperature}

\begin{figure}
 \begin{center}
 \includegraphics[angle=0,width=0.45\textwidth]{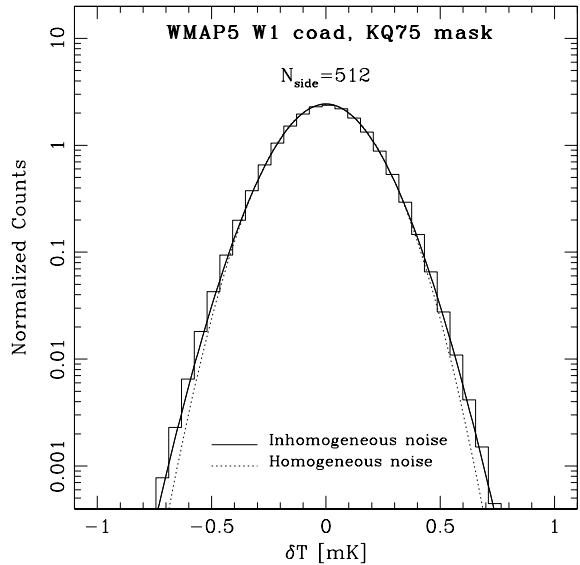}
 \caption{Distribution of temperature when $N_{\rm side} = 512$ for
          the WMAP5 W1 channel (histogram). Solid line shows the expected distribution (equation~\ref{pD}) 
          given the distribution of the noise.
          Dotted line is a Gaussian with the same rms as the data.
 \label{pDobs}}
 \end{center}
\end{figure}

The histogram in Figure~\ref{pDobs} shows the distribution of temperature in the 
data, for the W1 coadded channel. For homogeneous noise, the predicted distribution would be 
given by convolving the expected signal (in this case Gaussian with 
variance determined by smoothing the best-fit power spectrum on the 
scale of a pixel) with that of the noise (Gaussian with average
rms $\sigma_{\rm N}$).  
For inhomogeneous noise, the predicted distribution
(i.e. equation~\ref{pD}) is given by 
convolving the expected Gaussian signal with the noise of 
rms $\sigma_n$, and then weighting by the distribution of $\sigma_n$  
shown in Figure~\ref{psigma_hist}. 
The dotted line in Figure \ref{pDobs} shows a Gaussian with the same
rms as the data. This fit has shorter tails than the data.  
While this may be indicating that the signal is non-Gaussian 
(in fact, the sample skewness of the temperature distribution is
$-0.01216$ mK$^3$, and the sample excess kurtosis 0.12698 mK$^4$), 
the solid curve in Figure \ref{pDobs} shows the predicted distribution
when inhomogeneous noise effects are included: 
it provides substantially better agreement with the measurements.  

Nevertheless, the one-point distribution itself cannot rule out
a priori the presence of a non-Gaussian signal,
for reasons made explicit in Fang \& Pando (1997);
hence it is not a suitable statistics to investigate departures from Gaussianity.
This is also relevant for the detection of non-Gaussianity
claimed by Jeong \& Smoot (2007), where the one-point distribution was used.
We also note that their equation (9) is missing one integration over
the distribution of the noise (see again our equation \ref{pD} for comparison).

\begin{figure}
 \begin{center}
 \includegraphics[angle=0,width=0.49\textwidth]{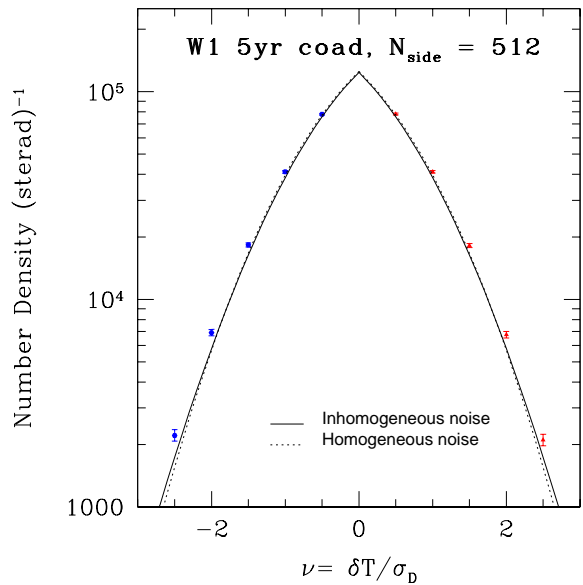}
 \caption{Number density of hot and cold pixels at
          $N_{\rm side}=512$, for the WMAP5 W1 channel. Points are
          measurements from the data.
          Dotted line is the theoretical prediction for homogeneous
          noise, and solid line includes inhomogeneity.
          Different temperature thresholds are considered, as
          explained in the main text.} 
 \label{number_density_512}
 \end{center}
\end{figure}

\subsection{Pixel Number Density}

\begin{figure*}
 \begin{center}
 \includegraphics[angle=0,width=0.496\textwidth]{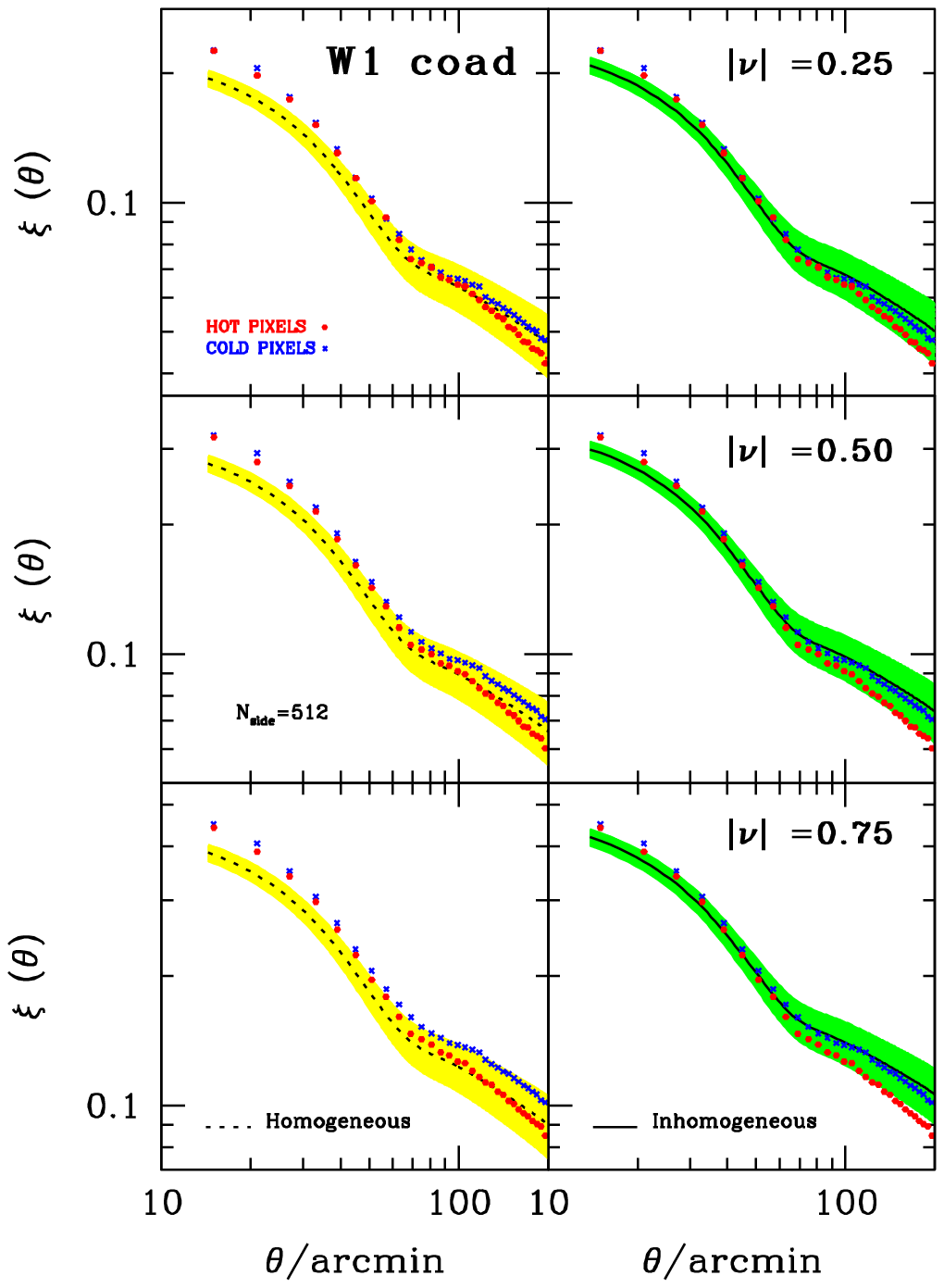}
 \includegraphics[angle=0,width=0.496\textwidth]{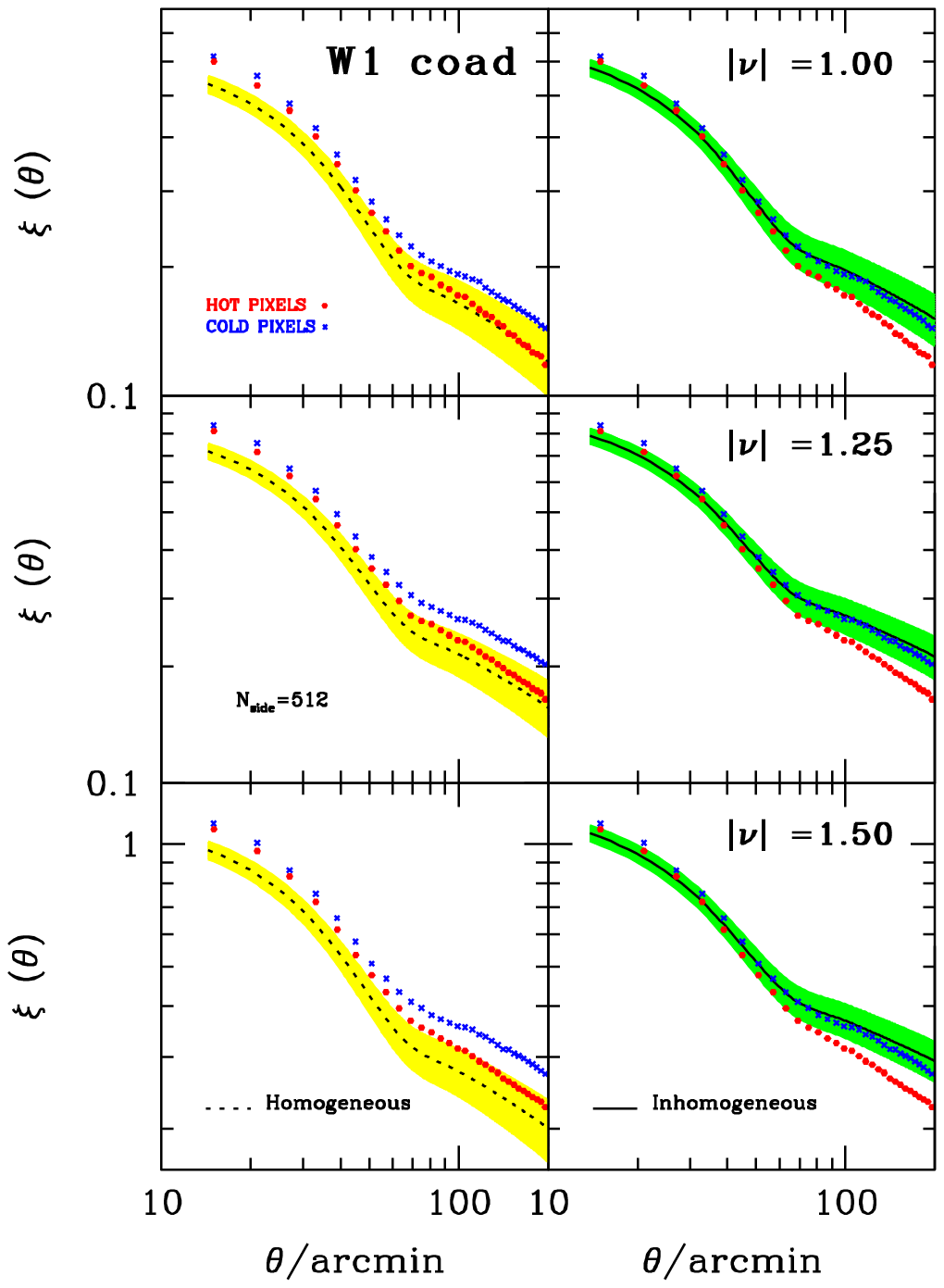}
 \caption{Unweighted correlation functions of pixels above threshold
          calculated from the WMAP5 W1 temperature 
          field, at $N_{\rm side}=512$.
          A variety of pixel-thresholds are considered, as indicated
          in the panels. Dotted curves in the left panels show the predictions associated with 
          Gaussian signal plus homogeneous noise, solid curves in the
          right panels show the case of inhomogeneous 
          noise. Points are measurements of the clustering of hot and cold pixels, at corresponding
          temperature thresholds. Shaded areas are the $1 \sigma$ optimal variance errors.
\label{xihi}}
\end{center}
\end{figure*}

\begin{figure*}
 \begin{center}
 \includegraphics[angle=0,width=0.496\textwidth]{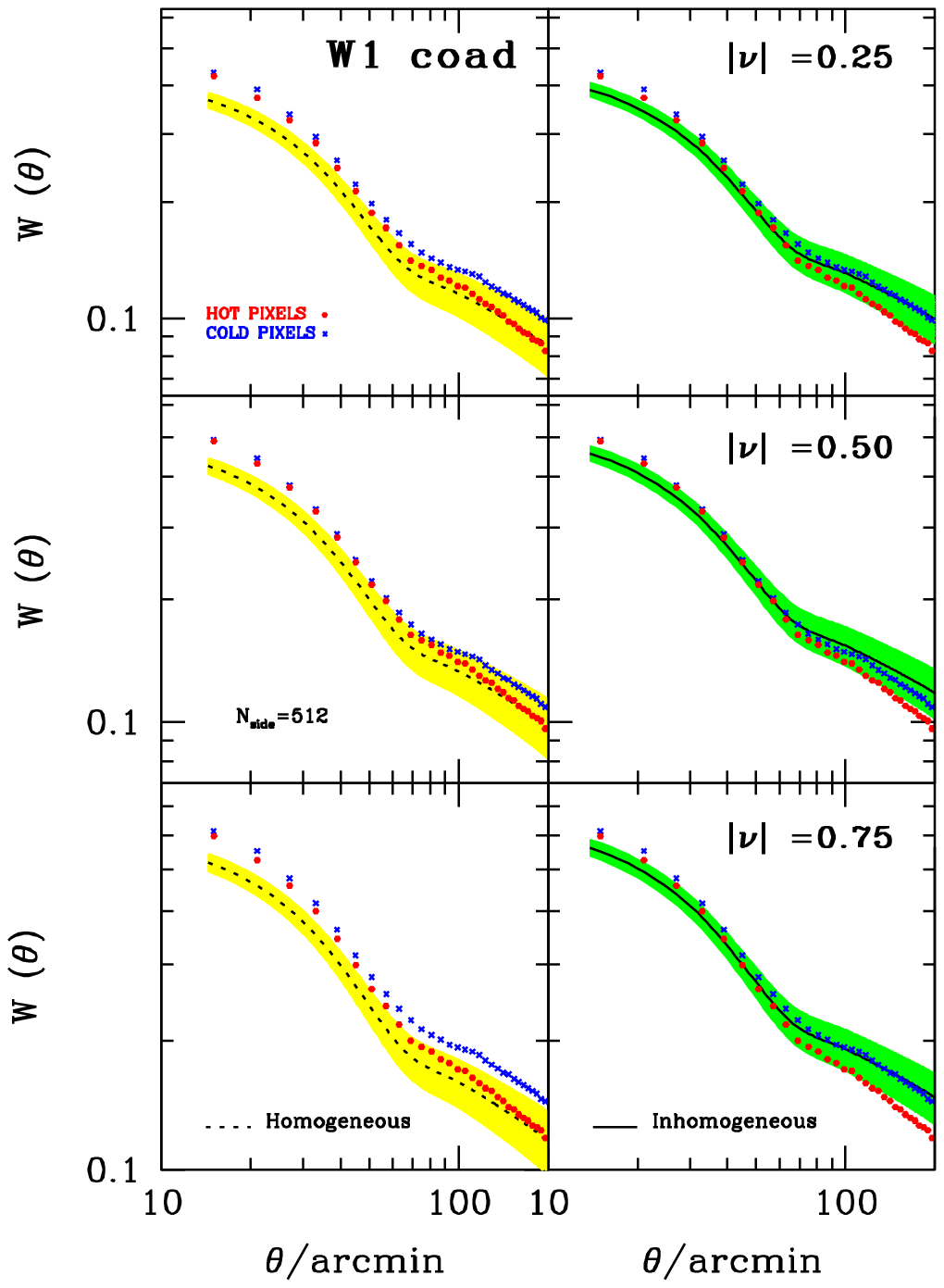}
 \includegraphics[angle=0,width=0.496\textwidth]{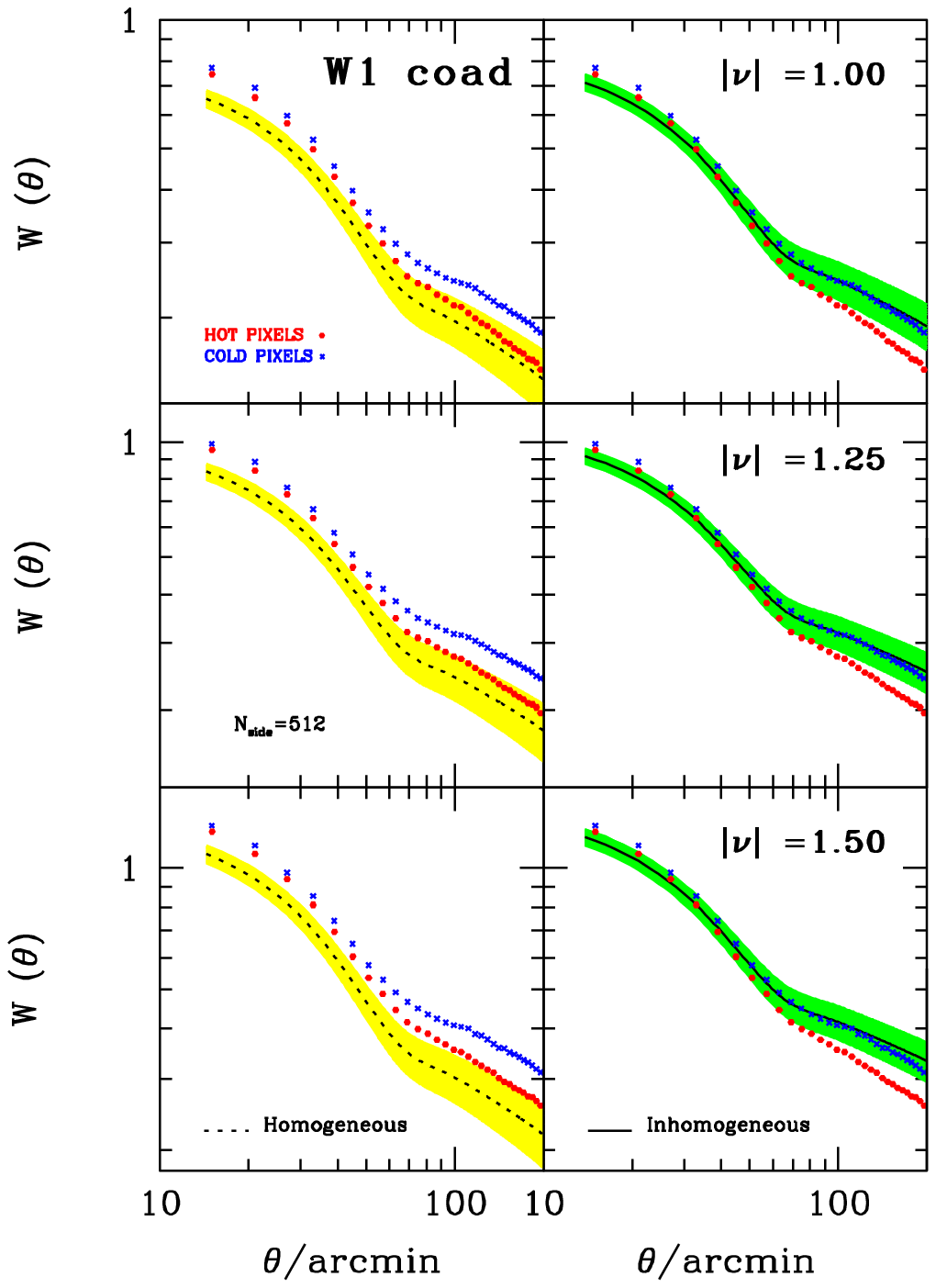}
 \caption{Same as Figure \ref{xihi}, but for the weighted correlation functions.} 
 \label{wihi}
 \end{center}
\end{figure*}

The number density of pixels above (below) a given temperature threshold is simply obtained by 
multiplying the fraction of pixels above (below) the threshold $D_{\rm
 t}$ (eq. \ref{fDt})
by the total number density after masking (i.e. the total number of pixels
after masking over the ``available'' or ``unmasked'' area).
If the noise were homogeneous, then this 
prediction is rather straightforward: equation (\ref{fDt}) simply reduces to
${\rm erfc} (\nu/\sqrt{2})/2$, where $\nu = D_{\rm t}/\sigma_{\rm D}$. 
In practice, due to the inhomogeneity of the noise, 
one needs to integrate (\ref{fDt}) numerically, using
the distributions $p(\sigma)$ measured from the data (Figure \ref{psigma_hist}). 
The solid lines in Figure \ref{number_density_512} are the results of this
integration, when accounting for inhomogeneous noise.
Dotted lines are the theoretical predictions which assume homogeneous
noise. Points in Figure \ref{number_density_512} show
measurements from the WMAP5 W1 channel, for 
pixels above or below a threshold, where $|\nu|=0.5,1.0,1.5,2.0,2.5$, respectively. 
Errorbars are from a Poisson analysis. 
While the solid lines slightly improve the fits, significant departures
still remain, especially at higher thresholds. 


\subsection{Two-point Statistics above Threshold}

We have measured the correlation function of pixels above and below a 
threshold temperature from the WMAP
five-year coadded maps, for a range of thresholds. In each case, 
we measured the signal in two ways: one in which all pixels 
are treated equally, and another in which the 
pixels are weighted by their temperature.
We adopt here the standard 
estimator $1+\xi(\theta) = DD(\theta)/RR(\theta)$ for the correlation
function (Jensen \& Szalay 1986), 
where $DD(\theta)$ and $RR(\theta)$ 
are the number of data and random pairs, respectively.
In particular,
the number of random pairs is computed by distributing random points
on a unit sphere, and then by applying the same
procedures (i.e. KQ75 masking and HEALPix coordinate assignment) as for the
data pipeline.
For the weighted correlation estimates, the number of data pairs,
$DD(\theta)$, is simply replaced by its temperature
weighted counterpart, $WW(\theta)$. We also tested the Landy \& Szalay
(1993) and the Hamilton (1993) estimators, and found no significant
differences in the calculations.

Figures~\ref{xihi} and~\ref{wihi} show an example of the results 
for pixels with $|\nu|=0.25$, $0.50$, $0.75$, $1.00$, $1.25$ and
$|\nu|=1.50$, respectively.  
In all panels, symbols show the measurements of hot and cold pixels
at the fixed threshold; errors are from a 
Poisson analysis. Dotted curves in all the left panels show the results based on 
homogeneous noise, solid curves in all the right panels account for the fact that 
the noise is inhomogeneous (equations~\ref{xitheta} and~\ref{Wtheta}).   
Shaded areas are derived as explained in Appendix A (see also Hern{\'a}ndez-Monteagudo et al. 2004), 
and represent the $1 \sigma$ optimal variance errors.

Accounting for inhomogeneity significantly improves the fit to the
data. However, discrepancies between the observed correlation functions and
the theoretical predictions still remain, especially for pixels at higher
thresholds, or at small angular scales. 
In particular, we note the interesting fact that hot and cold pixels cluster
differently within the same temperature thresholds. 
This feature is also present in all the other WMAP5 channels (see
Appendix B), but we find that our theory is always in good agreement with
the clustering of cold pixels in the Q and V frequency bands, and only in slight disagreement
with the hot patch clustering, especially for pixels at higher
thresholds.
Moreover, the fact that the WMAP5 data appear to be systematically stronger (weaker) at small (large)
angular scales than the theoretical predictions, even at lower
temperature thresholds (Figures \ref{xihi} and
\ref {wihi}), is particularly true in the W channel (see Figure
\ref{cf_512_nu_0.25_0.75_app} for comparison). We suggest some
possible interpretations in the discussion section and in Appendix B,
and present a more detailed investigation of this effect using realistic
non-Gaussian mock simulations in a forthcoming paper.

While the detected trend may be a unique 
signature of primordial non-Gaussianity (i.e. a primordial
non-Gaussianity would in fact enhance the clustering of the cold pixels and reduce
that of the hot ones), we study the effect of smoothing the map next,
and provide some other plausible explanations for these anomalies in the
final section.


\subsection{Effect of Smoothing}

We have investigated the effect of smoothing the WMAP five-year maps with a
Gaussian beam. In particular, we have tried different smoothing scales,
corresponding to a FWHM of $30'$, $45'$, $60'$
and $75'$, respectively. We have repeated the same analysis as for the
unsmoothed maps, and characterized the number density and the clustering
statistics of pixels above threshold.

\begin{figure*}
 \begin{center}
 \includegraphics[angle=0,width=0.78\textwidth]{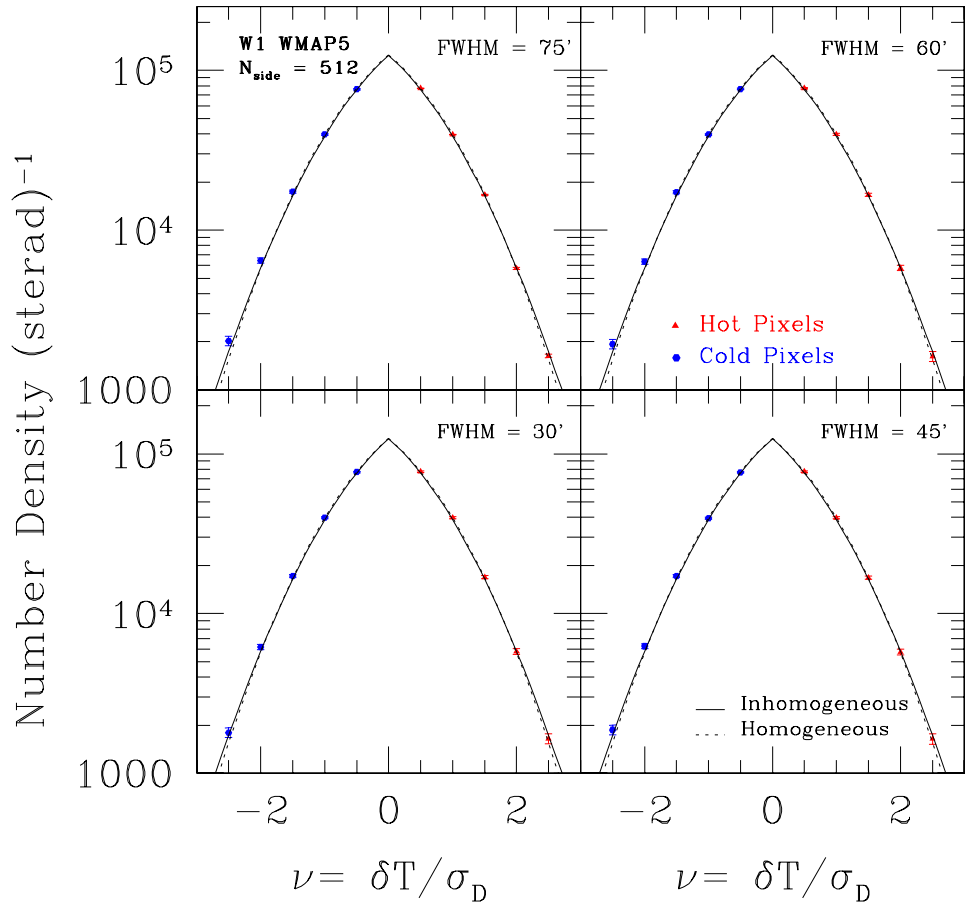}
 \caption{Number density of hot and cold pixels at
          $N_{\rm side}=512$, for the smoothed WMAP5 (W1 channel) maps.
          In each panel, points are measurements from data, dotted lines are the theoretical predictions for homogeneous
          noise, solid lines include inhomogeneity. From top to
          bottom and clockwise, the FWHM of the Gaussian beam is
          $75'$, $60'$, $45'$ and $30'$, respectively.}
 \label{number_density_512_smooth}
 \end{center}
\end{figure*}

Figure \ref{number_density_512_smooth} summarizes the results for the number
density (W1 channel), after smoothing with different Gaussian beams. The number of pixels
above or below the considered temperature threshold is reduced (a function of the
smoothing scale), and measurements from the data and theoretical predictions
appear now in agreement within about $1 \sigma$ at the threshold
levels considered.    

\begin{figure*}
 \begin{center}
 \includegraphics[angle=0,width=0.48\textwidth]{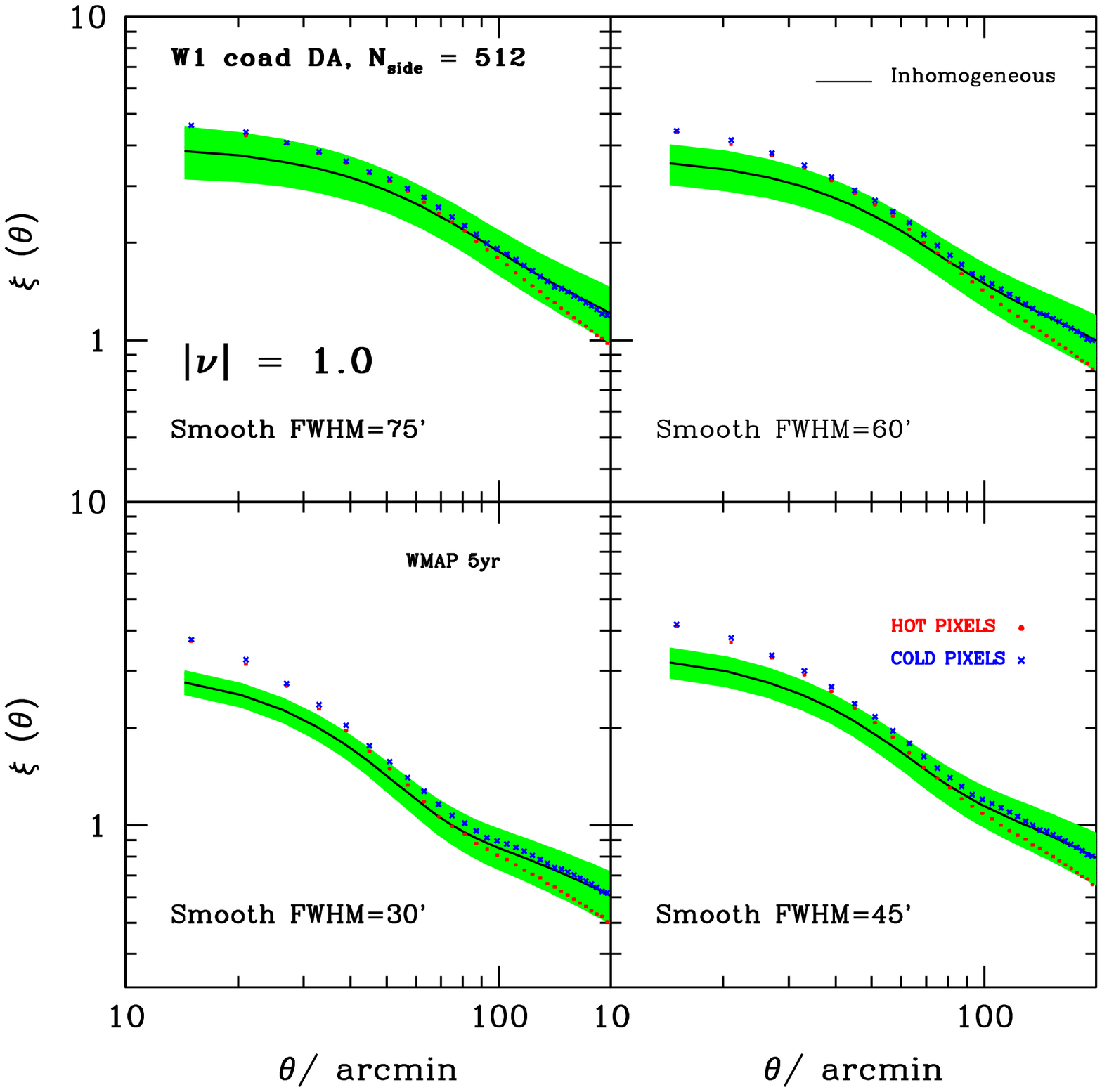}
 \includegraphics[angle=0,width=0.48\textwidth]{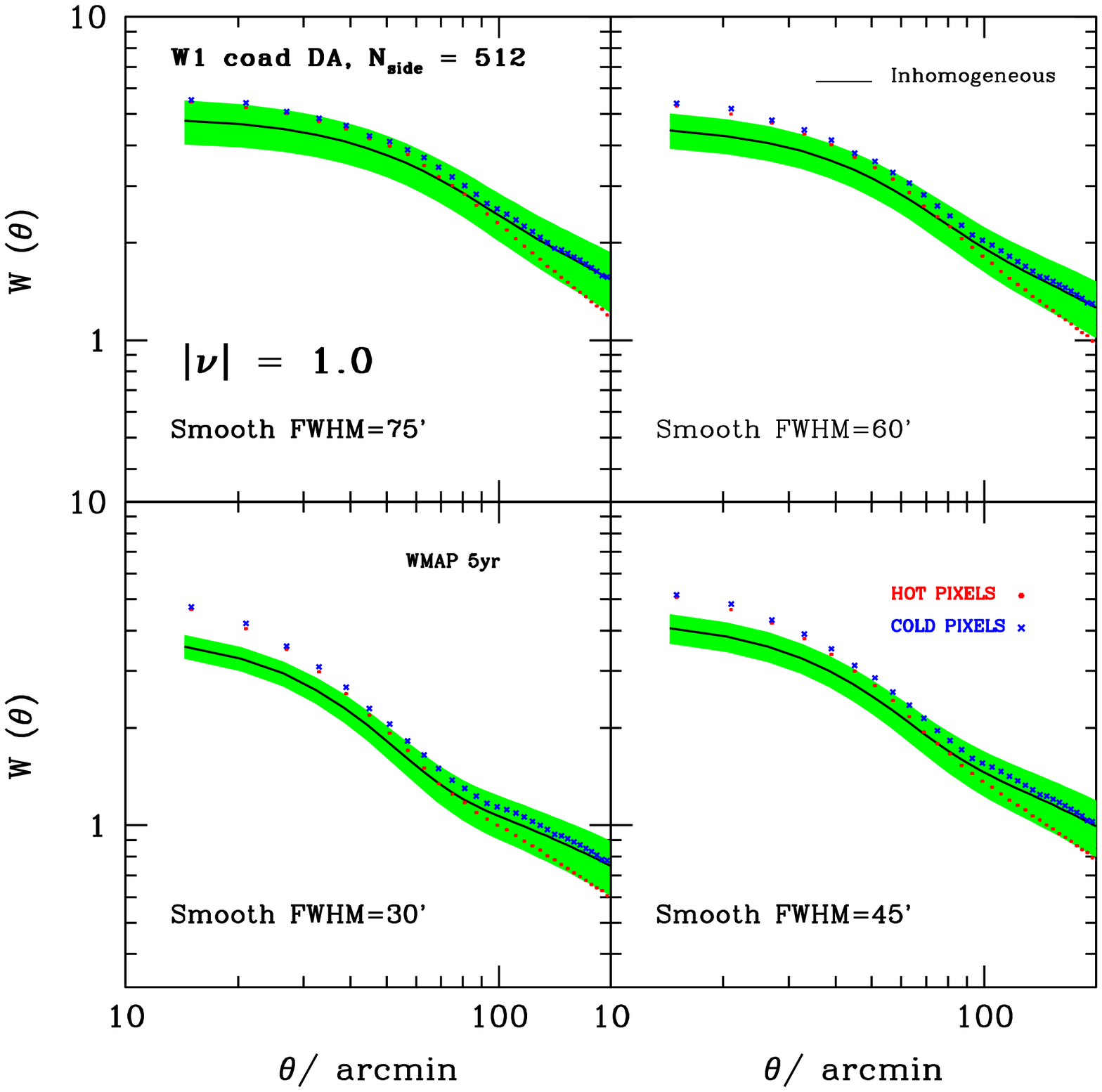}
 \caption{Unweighted [left panel] and weighted [right panel] correlation functions from the WMAP5 pixel-pixel temperature 
          field smoothed at different scales, at $N_{\rm side}=512$,
          for the W1 channel.
          The pixel threshold considered is $|\nu|=1.0$.
          In each panel, from top to
          bottom and clockwise, the FWHM of the Gaussian beam is
          respectively $75'$, $60'$, $45'$ and $30'$.
          Solid curves show the case of inhomogeneous 
          noise. Points are measurements of the clustering of hot and cold pixels, at corresponding
          temperature thresholds. Shaded areas are the $1 \sigma$ optimal variance errors.}
 \label{clustering_smooth_512}
 \end{center}
\end{figure*}

Finally, we have characterized the clustering statistics
and showed in Figure \ref{clustering_smooth_512} the case of inhomogeneous
noise, for an arbitrary choice of the threshold ($|\nu|=1.0$).
The Gaussian smoothing is included in our theory via equations
(\ref{ctheta_eq}) and (\ref{ctheta_noise}), in addition
to the change of the noise variance distribution. 
After smoothing the map with a Gaussian beam, 
the pixel noise may be no longer independent among different pixels (i.e. $\langle n_1 n_2 \rangle \ne 0$).
However, accounting for this contribution (equation \ref{n1n2_cov}) 
resulted in no significant difference, even at small angular scales.
In fact, the average variance of the noise in the W1 channel is 0.0213
mK$^2$, which is about 0.32$\%$ of the typical variance of the 
corresponding cosmological signal. Since the zero-level amplitude of the noise is
already very low with respect to that of the signal,    
the mean amplitude of the correlated noise will be also very low.
In addition to that, note that the WMAP mission has been designed to
have uncorrelated pixel noise, and if such
correlations arise when smoothing the maps, their amplitude is
expected to drop quickly with the smoothing scale.

When the smoothing scale is $75'$, we find that the tension between
data and theory is alleviated. 

\section{Discussion}

The observed distribution of temperatures in WMAP5 pixels is slightly 
non-Gaussian (Figure~\ref{pDobs}). This departure from Gaussianity 
is not unexpected if the intrinsic signal is Gaussian but the noise 
distribution is inhomogeneous white noise (Figures~\ref{psigma_map}
and \ref{psigma_hist}).  
In fact, when accounting for inhomogeneous noise (Section \ref{noise_model}), we find that the one-point statistics
is substantially in better agreement with the measurements.  
However, the predicted dependence of clustering on pixel height is not in good agreement 
with the data, even after accounting for the inhomogeneity of 
the noise (Figures \ref{xihi}, \ref{wihi}).  
Although we found that an appropriate smoothing scale would be able to
alleviate the discrepancy between data and measurements (Figures \ref{number_density_512_smooth} and \ref{clustering_smooth_512}), 
other possible reasons for these discrepancies are: \\

\textit{Masking procedure}. In our analysis we adopted the KQ75
  mask, which allows for selective exclusions of bright portions of the
  sky ($28 \%$ pixel cut). Although this is rather a drastic cut and
  the mask was significantly improved with the WMAP five-year data release
  (Hill et al. 2008), it may not completely account for
  all the Galactic foreground effects. 
  However, cutting out all the Galactic plane ($|b| \le 30^{ \circ }$ strip) 
  and repeating our analysis resulted in no significant change.
  Edge effects due to pixels which lie very close to the mask could also affect our
  analysis. \\
\textit{Cold spot contamination}. 
  Inoue \& Silk (2007) suggest that the presence of low density 
  regions in the southern Galactic cap could account for our 
  anomalies, but restricting our analysis to $b \ge 30^{\circ}$, 
  the northern Galactic cap, resulted in no significant change.  \\
\textit{Contamination by point sources}.  
  Lopez-Caniego et al. (2007) detected 98 new sources (i.e. $26 \%$) 
  not present in the WMAP three-year catalog. At higher frequencies WMAP 
  estimates neglect the deviations of the point spread function
  from a Gaussian shape, and carry out a blind search for sources. 
  With the WMAP five-year data release, the mask for   
  point sources has been significantly improved (Wright et al. 2008),
  and other studies (Gonzalez-Nuevo
  et al. 2008; Chen \& Wright 2008) updated the point source catalog.
  However, Massardi et al. (2008) recently provided a new version of
  the catalog based on WMAP5 data and found new point sources 
  (484 sources detected), carrying
  a complementary blind and non-blind approach.    
  This may be important to our analysis, since even a very low level 
  contamination in the maps can produce spurious
  non-Gaussianities. In fact, we find some discrepancies at small
  angular scales (around 20') especially in the W band, and their origin may be associated
  with extragalactic dust emission (one type of point-source contamination), which peaks
  at high frequencies.     
  In a forthcoming study, we will be addressing the effect of contamination   
  induced by point sources in more depth, using an updated version of
  the source catalog and mock simulations. \\ 
\textit{Foreground subtraction contamination}. 
  Uncertainties in the external Foreground Template Model used for the 
  foreground subtractions (Gold et al. 2008)
  may introduce anomalies at the percentage level. 
  The template itself has noise, which may be correlated at small separations. 
  This is a delicate issue, since galactic foregrounds are non-Gaussian and
  anisotropic, and even low level contamination
  in the maps can produce detectable non-Gaussianities (Park et
  al. 2002; Naselsky et
  al. 2005; Kim, Naselsky \& Christensen 2008). On the other side, it
  is also worth noticing that recently Vio \& Andreani (2008) showed that the benefits of
  using more sophisticated methods for foreground cleaning, such as the
  Harmonic Internal Linear Combination, are overestimated. We are
  also addressing the foreground subtraction systematics in a forthcoming
  study. \\
\textit{WMAP beam, window function, absolute calibration and pixel-noise uncertainties}.
  Window functions were computed from the symmetrized beam 
  profiles following the Hermite method in Page et al. (2003). 
  A typical WMAP window function has an uncertainty of $2-3\%$
  (Hill et al. 2008), 
  and these uncertainties add in quadrature in the cosmological 
  analysis. Absolute calibration uncertainties in the five-year 
  WMAP data are estimated to be $0.5\%$ (Hinshaw et al. 2008). 
  Recent studies (Groeneboom et al. 2009; Kathrine Wehus et al. 2009)
  have also pointed out that the noise levels of these maps are
  underestimated, and that there are some problems with the standard
  WMAP transfer functions as well.    
  It may be that these facts play an important rule in our study (see also
  Colombo et al. 2008).\\
\textit{Secondary anisotropies and phase transitions in the early universe}.
  Spurious non-Gaussianities 
  could arise from secondary anisotropies, such as gravitational
  lensing, Sunyaev-Zel'dovich effect or Sachs-Wolfe effects (Babich \&
  Pierpaoli 2008; Carbone et al. 2008). Phase transitions in the early
  universe may also introduce a new source of non-Gaussianity
  (Silvestri \& Trodden 2008). All these effects may be difficult to
  disentangle from a pure primordial non-Gaussian signal, even after a
  clear detection of primordial non-Gaussianity.
\textit{Real non-Gaussian signatures at small scales}.
  A better understanding of all the previous points is necessary before 
  we can claim that the signal we see is due to primordial 
  non-Gaussianity. However, we found an interesting difference in the
  clustering of hot
  and cold pixels within the same temperature threshold level; 
  this fact may be a unique signature of primordial
  non-Gaussianity. We present a detailed investigation of this feature
  in a forthcoming study.    
  We finally note that, as this work was being refereed, Hou et al. (2009) presented a frequentist
  analysis of the correlation functions of the local extrema, and also found inconsistency
  with Gaussian simulations plus differences in the clustering of hot and cold
  peaks -- although comparison with their work is not direct since
  they considered bigger angular scales.
 
While seeking for primordial non-Gaussianity is at the moment
a new frontier in cosmology, ongoing efforts are currently devoted to
the characterization of non-Gaussian confusion effects, to
reliable theoretical predictions of non-Gaussianity from models (see
for example Boyle \& Steinhardt 2008; Fergusson \& Shellard 2008;
Munshi \& Heavens 2009), 
till the extraction of information from data 
(i.e. Raeth et al. 2008; Gong et al. 2009) or the search for observational signatures of
primordial non-Gaussianity imprinted in the large-scale structure of the universe.
Even a small degree of primordial non-Gaussianity can be a crucial probe of the
inflationary dynamics or alternative universe models, hence studies of
non-Gaussianity may eventually become a powerful and solid probe of
ultra-high energy physics and inflation.

We note that our model for the effects of inhomogeneous noise 
may be useful in other studies (see for instance Yu \& Lu 2008).
Extending our formalism for inhomogeneous noise to peak rather than 
pixel statistics (for example Heavens \& Sheth 1999; Heavens \& Gupta 2001) 
is more complicated; this will complement numerical Monte-Carlo 
analyses of this problem (Larson \& Wandelt 2004; Tojeiro et al. 2006;
Ayaita et al. 2009; Hou et al. 2009), 
and is the subject of work in progress.  
   


\section*{Acknowledgments}

We thank an anonymous referee for helpful comments and suggestions.
GR thanks Fernando Atrio-Barandela, Pravabati Chingangbam, Jacek Guzik,
Mike Jarvis and Licia Verde for many useful discussions.  
GR is grateful to the organizers of the 24th Texas Symposium on Relativistic Astrophysics for their kind
hospitality in Vancouver, during which time he had interesting conversations with
Eiichiro Komatsu and the final stage of this work was completed. GR
would also like to thank KITPC and in particular Robert Brandenberger 
for hospitality in Beijing during
the 2009 workshop ``Connecting Fundamental Physics with Observations''. 
CBP acknowledges the support of the Korea Science and Engineering
Foundation (KOSEF) through the Astrophysical Research Center for the
Structure and Evolution of the Cosmos (ARCSEC).
We acknowledge the use of the \textit{L}egacy \textit{A}rchive for 
\textit{M}icrowave \textit{B}ackground \textit{D}ata \textit{A}nalysis 
(\textit{LAMBDA}), support for which is provided by the National 
Aeronautics and Space Administration (NASA) Office of Space Science. 
Some of the results in this paper have been derived using the 
\textit{HEALPix} package (G\'orski et al. 1999). 
This work was partly supported by NASA grant NNG05GK81G, and 
partly by the Project of Knowledge Innovation Program (PKIP) of Chinese Academy of Sciences, Grant No. KJCX2.YW.W10.




\appendix

\section{Accounting for Finite Binsize Effects}

We estimate the uncertainties in the correlation functions above threshold from
the optimal variance limit, containing cosmic variance, instrumental
noise, and finite binsize effects.
 
In more detail, the ultimate accuracy with which the CMB power spectrum can
be determined at each $l$ is given by (Knox 1995):
\begin{equation}
\sigma_{\rm OV}^2 (l) = \sqrt{{2 \over (2\,l+1)f_{\rm sky}}}\, \Big [
    C_{\rm l} + { 4\,
    \pi\, \sigma_{\rm N}^2 \over N \, W_{\rm l}^{\rm WMAP}} \Big ] \equiv \Delta
    C_{\rm l}\,
\label{opt_var}
\end{equation}
\noindent where $N$ is the number of pixels, $\sigma_{\rm N}$ the average
pixel noise, $W_{\rm l}^{\rm WMAP}$ the WMAP window function, and $f_{\rm sky}$ the fraction of
the sky covered by the experiment. 
The uncertainty in the angular correlation function for
narrow bins in $\theta$ is then (Hern{\'a}ndez-Monteagudo et al. 2004):
\begin{eqnarray}
\sigma^2_{\rm C(\theta)} &\equiv& \Delta C (\theta) =
  \Big\{ \sum_{\rm l} \Big | {\partial C(\theta) \over \partial
  C_{\rm l}}  \Big |^2 \, \Delta C_{\rm l}^2 \Big\}^{1/2} \nonumber \\
  &=& \Big\{ \sum_{\rm l} {(2l+1)
  \over 8 \pi^2 f_{\rm sky}} |P_{\rm l}^0(\cos \theta)|^2(W^{\rm WMAP}
  W^{\rm smooth})^2 \times \nonumber \\
  & & \times \Big[{2 \pi C_{\rm l}^{\star} \over l(l+1)} +
  {\Omega_{\rm pix} \sigma_{\rm N}^2 \over W^{\rm WMAP}}
  \Big]^2\Big\}^{1/2}
\label{delta_C_theta}
\end{eqnarray}
\noindent where $C_{\rm l}^{\star} = l(l+1) C_{\rm l}/2 \pi$ and
$W^{\rm smooth}$ is the
additional smoothing due to finite pixel size, 
mask influence and an optional Gaussian beam.

However, in practice $C(\theta)$ is not measured at a
point, but smeared out over a region of size $\Delta \theta$.
In the limit of large $l$ and small $\theta$, $P_{\rm l}(\cos \theta)
\sim J_0[(l+1/2)\theta]$ (Bond \& Efstathiou 1987). If the bin size is
not infinitesimal, one must replace:
\begin{eqnarray}
J_0[(l+1/2) \theta] &\rightarrow& 
{2 \over (t_{\rm max}^2 -t_{\rm min}^2)} \int_{t_{\rm min}}^{t_{\rm
    max}} J_0 [(l+1/2)t]~t~{\rm d}t \nonumber \\ &=&   { 2[t_{\rm max} J_{1}(k t_{\rm max}) - t_{\rm min} J_1
  (k t_{\rm min})] \over k (t_{\rm max}^2 -t_{\rm
    min}^2)}  
\label{J0}
\end{eqnarray} 
\noindent where $t_{\rm min} = \theta - \Delta \theta/2$ and  $t_{\rm max} = \theta + \Delta \theta/2$. 
This expression should then substitute $P_{\rm l}(\cos \theta) \sim J_0[(l+1/2)\theta]$
in (\ref{delta_C_theta}), to give $\sigma^2_{\rm C_{\Delta
\theta}(\theta)}=  \Delta C_{\Delta \theta} (\theta)$. 
The uncertainties in the correlation function above threshold are finally
derived from:
\begin{equation}
\sigma^2_{\xi_{\nu}(\theta)} = 
\Delta \xi_{\nu} (\theta) = \Big | {\partial \xi_{\nu} (\theta) \over \partial
   C(\theta)}  \Big | \, \Delta C_{\Delta \theta} (\theta)   
\label{delta_xi_theta}~.
\end{equation}


\section{Results For All the Other WMAP five-year Differencing Assemblies}

\begin{figure*}
\begin{center}
\includegraphics[angle=0,width=0.48\textwidth]{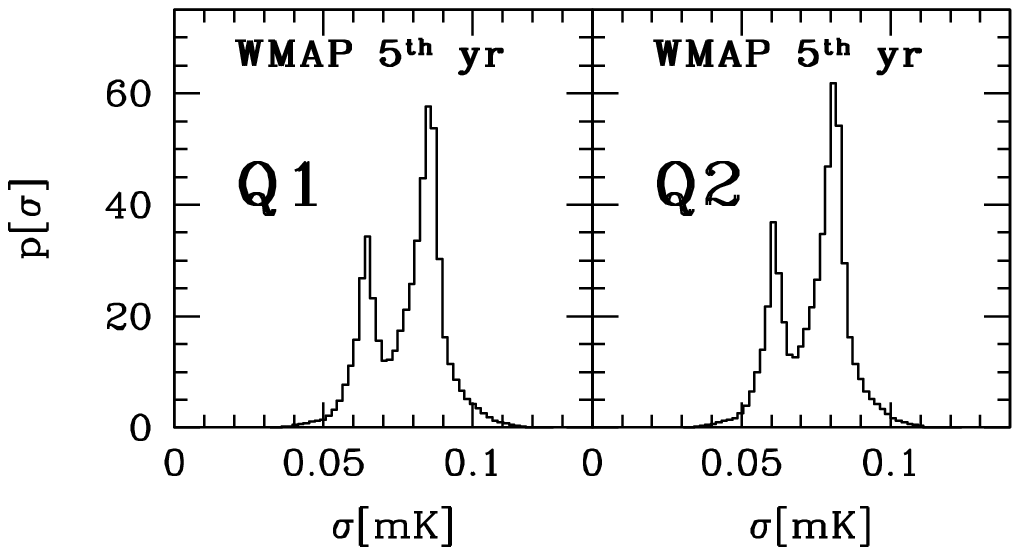}
\includegraphics[angle=0,width=0.48\textwidth]{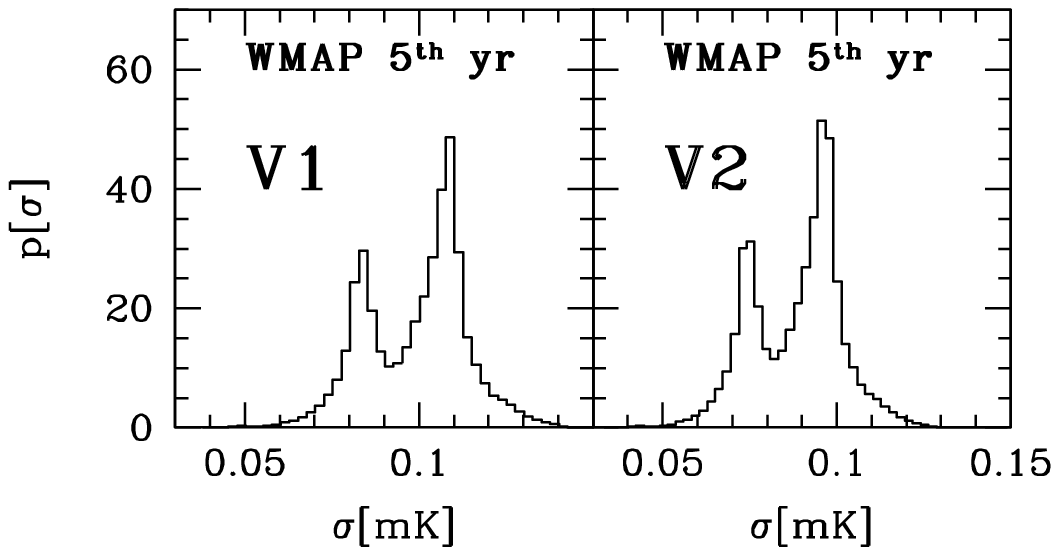}
\includegraphics[angle=0,width=0.70\textwidth]{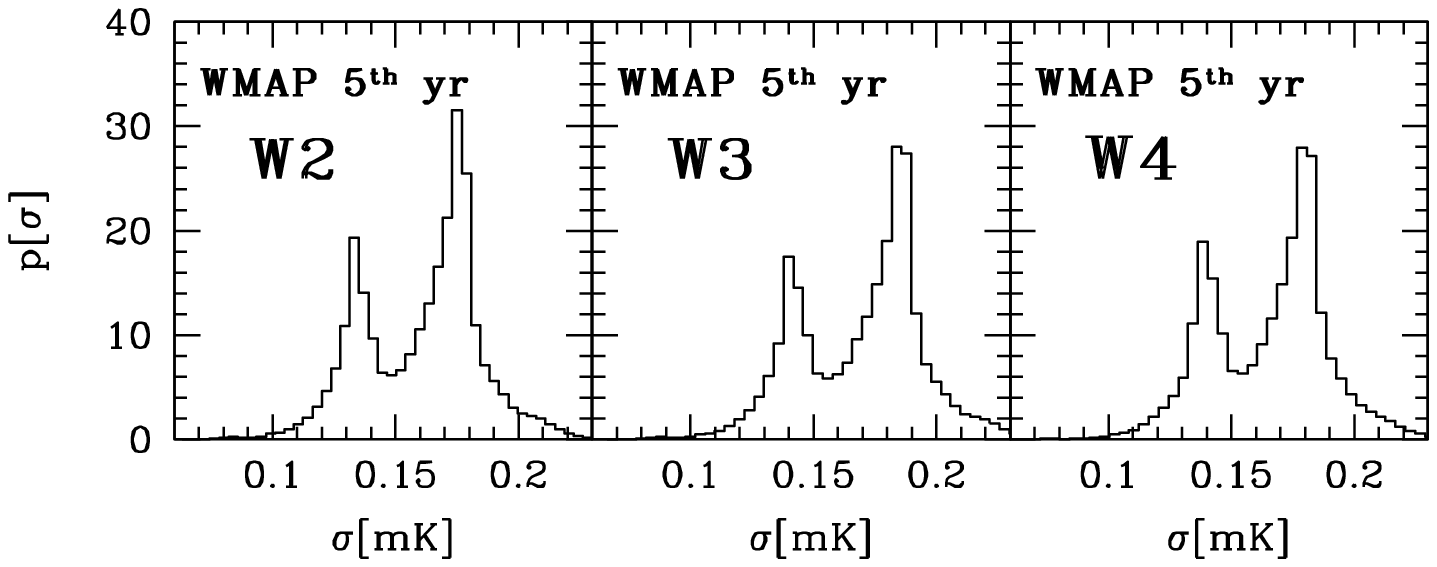}
\caption{Distributions of the rms noise-per-pixel at
         $N_{\rm side}=512$ for the five-year coadded WMAP 
         data, after application of the KQ75 mask. From top to bottom and from left to
         right, the various differencing assemblies are respectively
         Q1 and Q2 (the 41 GHz channel), V1 and V2 (the 61 GHz
         channel), and W2, W3, W4 (the 94 GHz channel).}
\label{psigma_map_app}
\end{center}
\end{figure*}

\begin{figure*}
 \begin{center}
 \includegraphics[angle=0,width=0.48\textwidth]{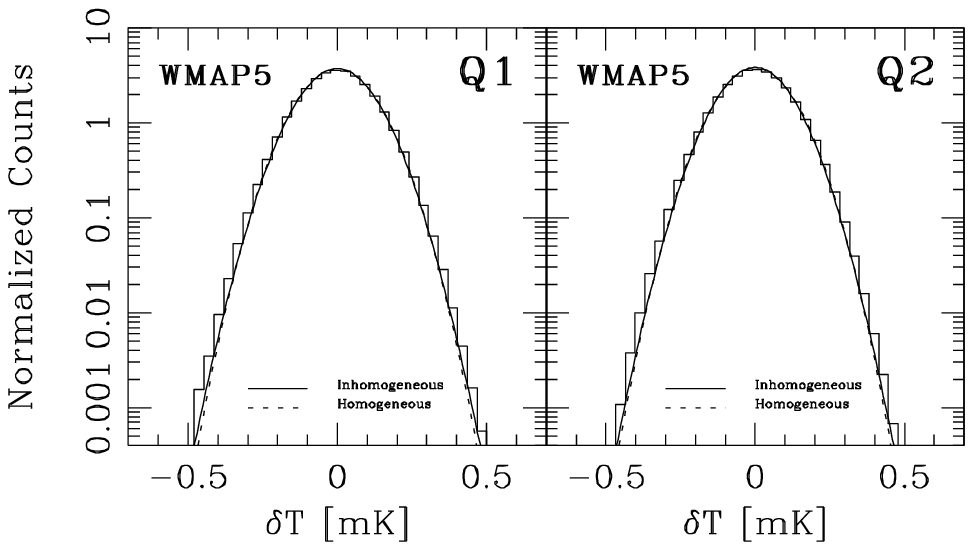}
 \includegraphics[angle=0,width=0.48\textwidth]{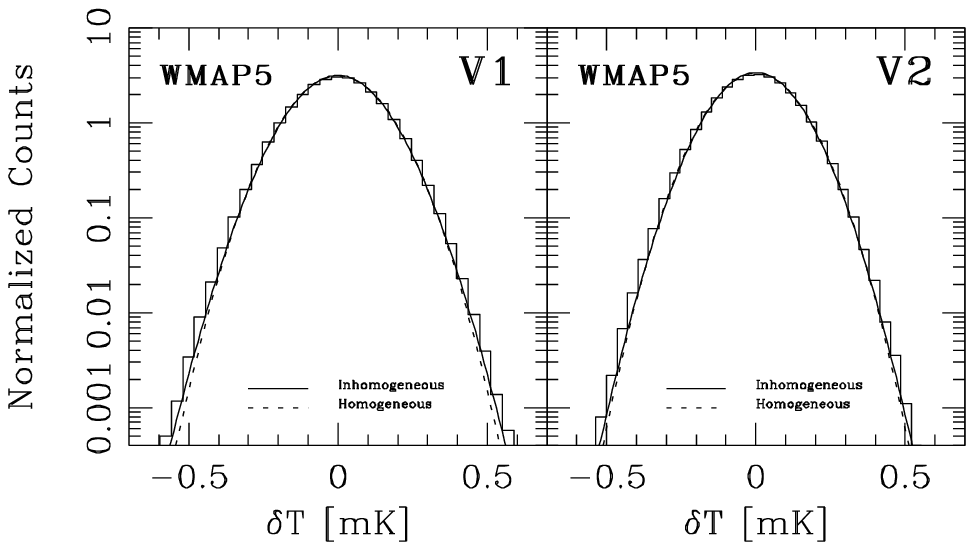}
 \includegraphics[angle=0,width=0.73\textwidth]{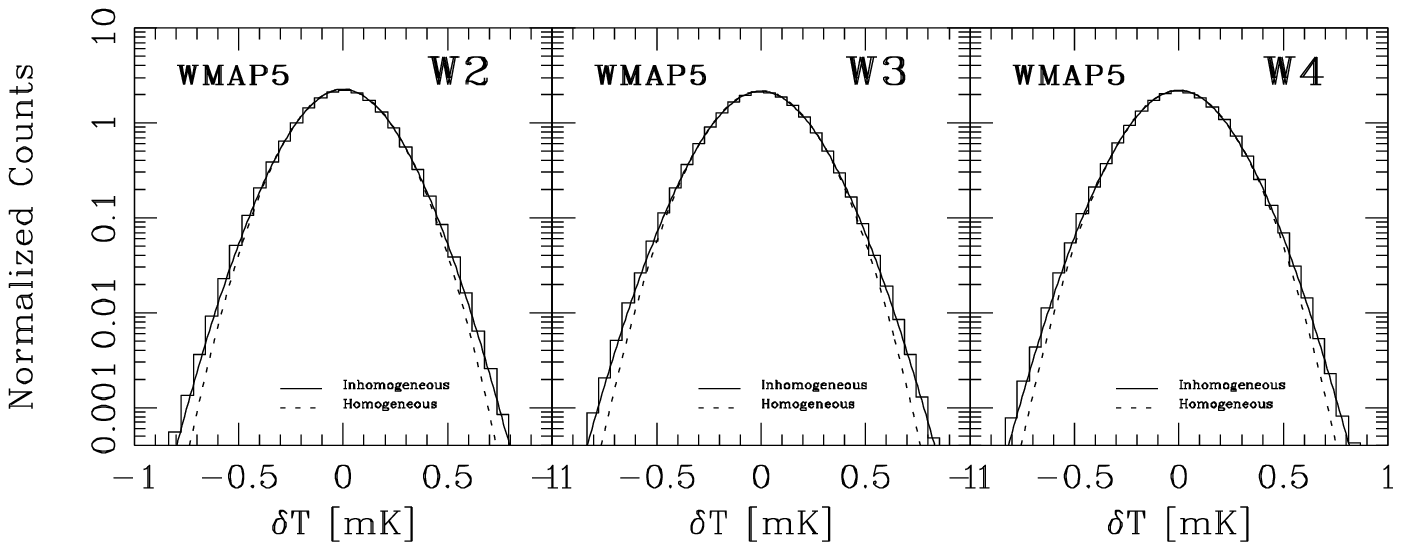}
 \caption{Distribution of temperatures when $N_{\rm side} = 512$ (histograms).
          Solid lines in all panels show the expected distributions (equation~\ref{pD}) 
          given the corresponding distributions of the noise.
          Dotted lines are Gaussians, with the same rms as the data. From top to bottom and from left to
          right, the various DA's are respectively Q1, Q2, V1, V2, W2, W3, W4.}
 \label{pDobs_app}
 \end{center}
\end{figure*}
\begin{figure*}
 \begin{center}
 \includegraphics[angle=0,width=0.65\textwidth]{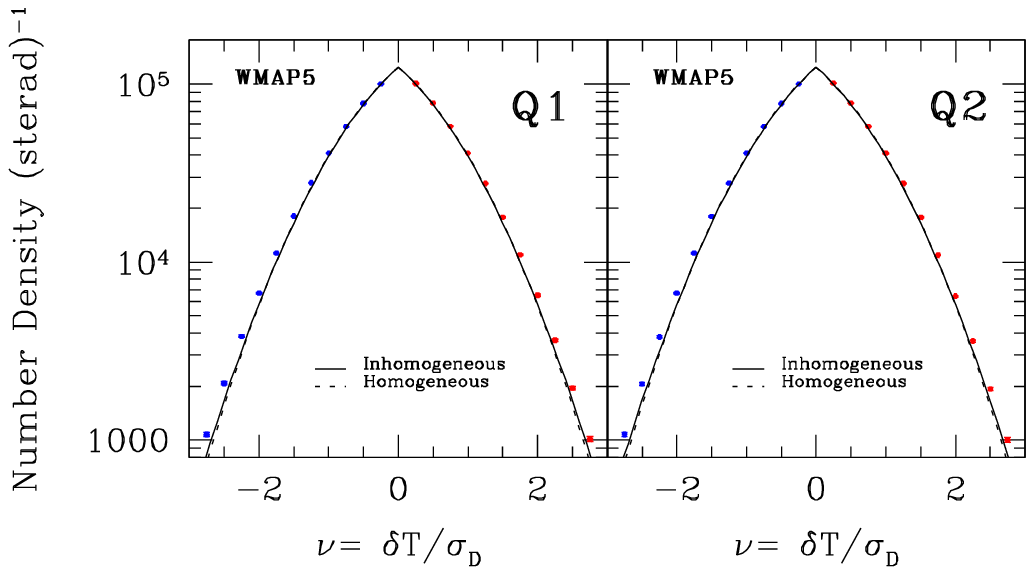}
 \includegraphics[angle=0,width=0.65\textwidth]{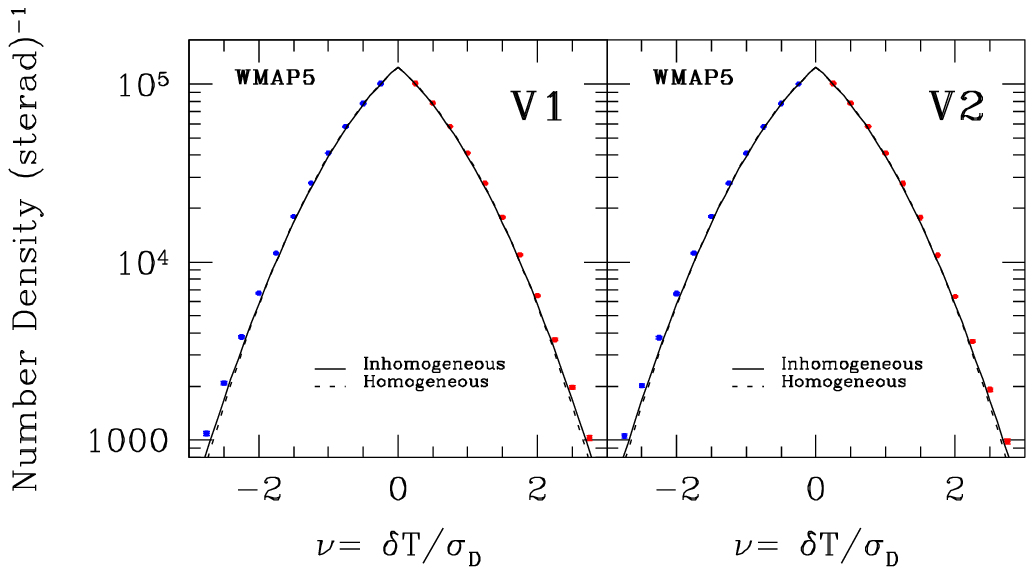}
 \includegraphics[angle=0,width=0.90\textwidth]{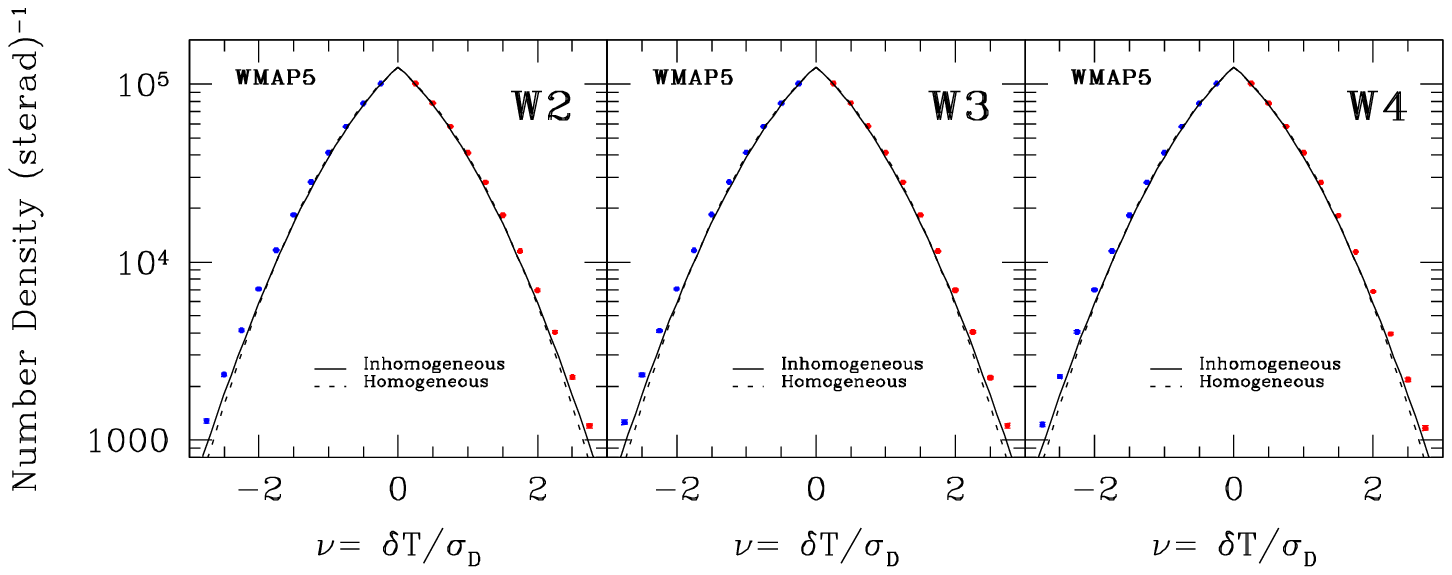}
 \caption{Number densities of hot and cold pixels at
          $N_{\rm side}=512$, for the WMAP5 individual channels. Points are
          measurements from the data.
          Dotted lines in all the panels are theoretical predictions for homogeneous
          noise, solid lines include inhomogeneity.
          From top to bottom and from left to
          right, the various DA's are respectively Q1, Q2, V1, V2, W2, W3, W4.}
 \label{number_density_512_app}
 \end{center}
\end{figure*}
\begin{figure*}
 \begin{center}
 \includegraphics[angle=0,width=0.48\textwidth]{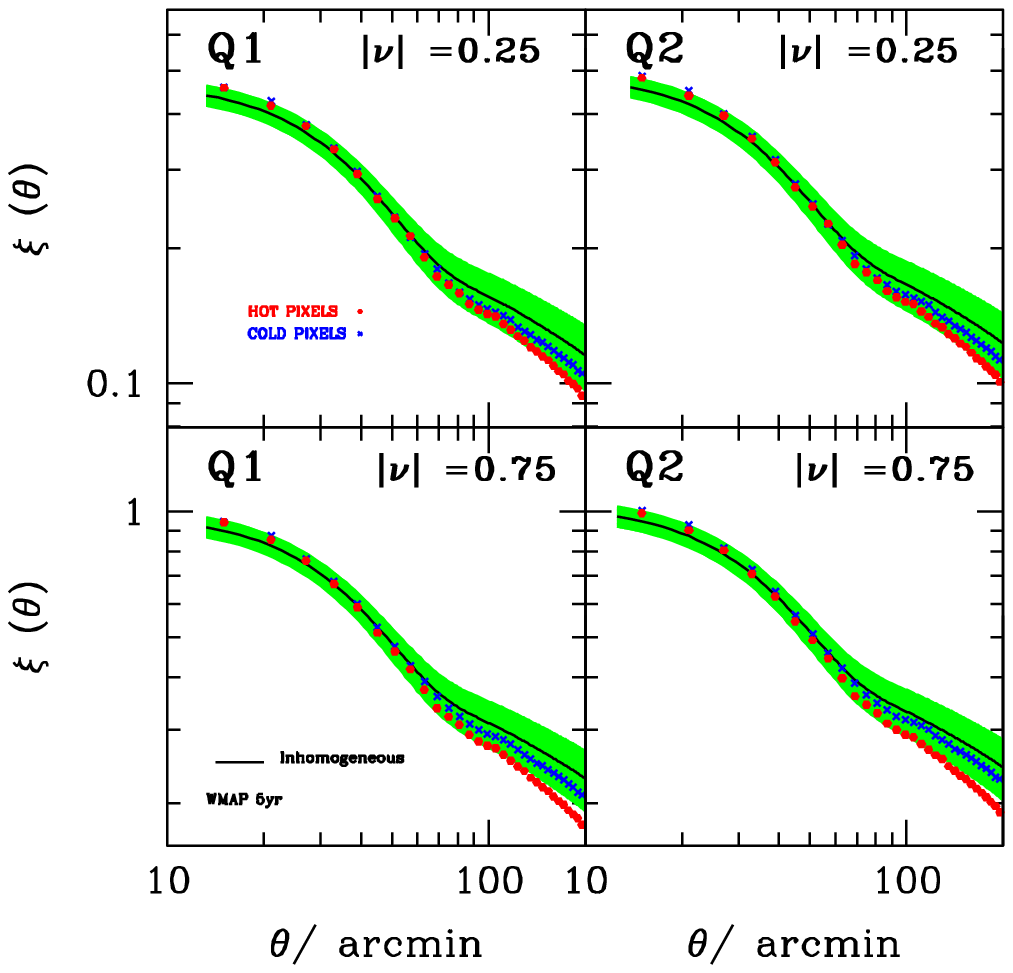}
 \includegraphics[angle=0,width=0.48\textwidth]{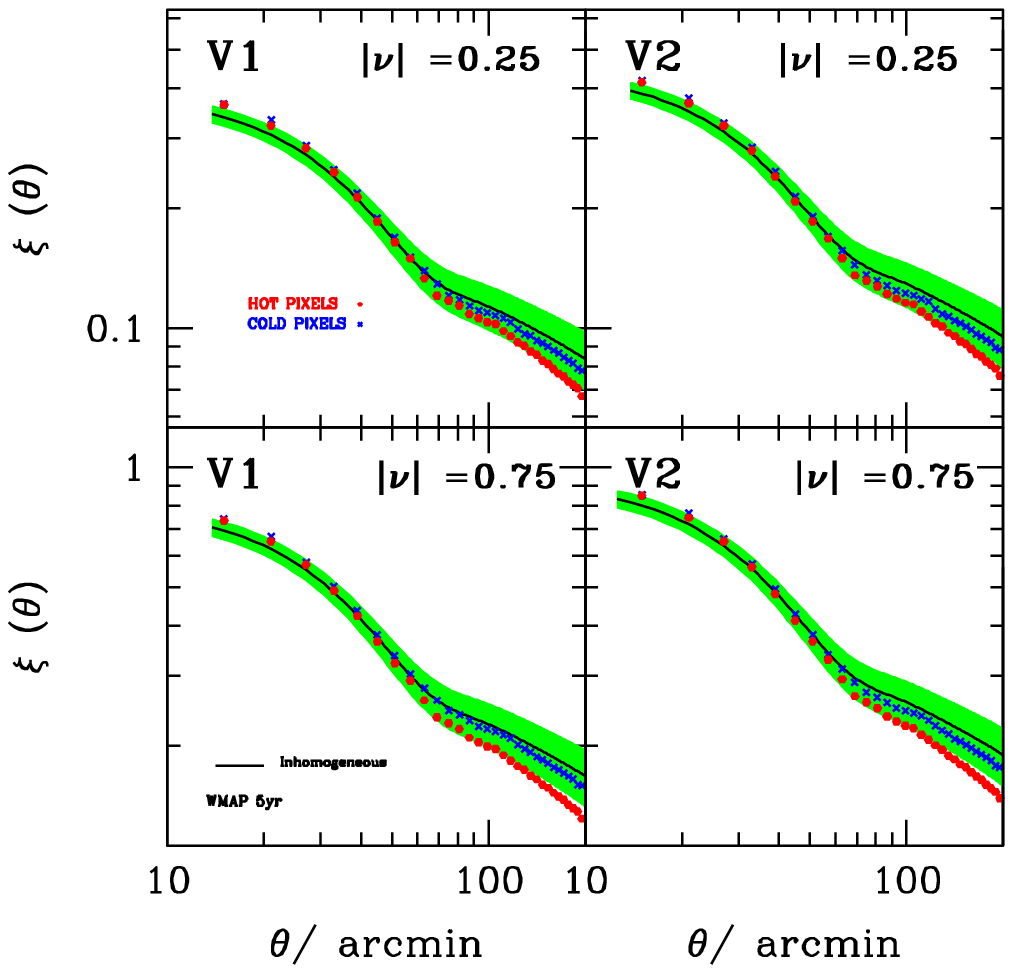}
 \includegraphics[angle=0,width=0.80\textwidth]{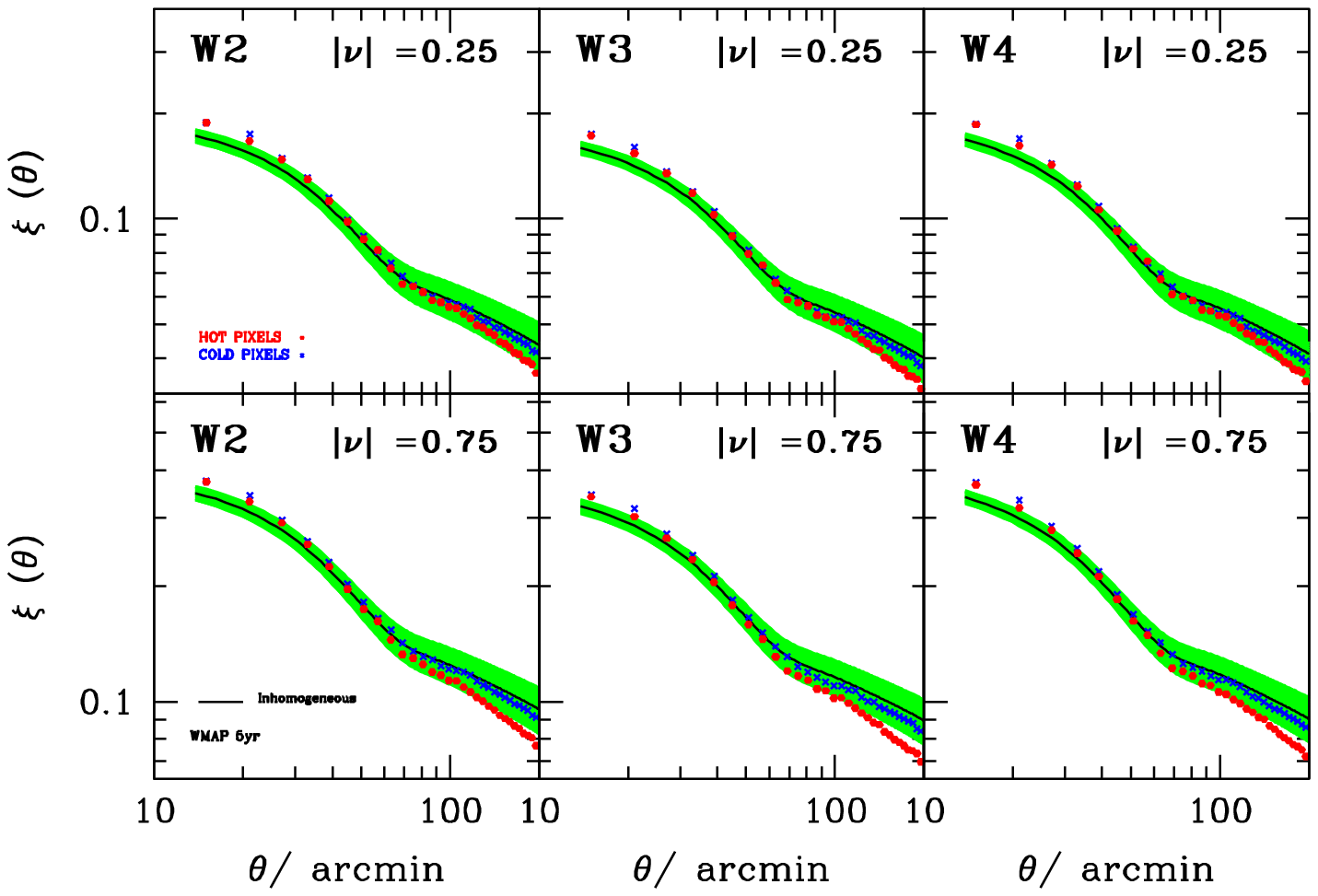}
 \caption{Examples of unweighted correlation functions from the WMAP5
 pixel-pixel temperature field, at $N_{\rm side}=512$. Two
 pixel-thresholds are considered ($|\nu|=0.25$ and $|\nu|=0.75$), as
 indicated in the panels. Solid curves in all panels show the
 predictions associated with Gaussian signal plus inhomogeneous
 noise. Points are measurements of the clustering of hot and cold
 pixels, at corresponding temperature thresholds. Shaded areas are the
 $1\sigma$ optimal variance errors. From top to bottom and from left to
 right, the various DA's are respectively Q1, Q2, V1, V2, W2, W3, W4.}
 \label{cf_512_nu_0.25_0.75_app}
 \end{center}
\label{lastpage}
\end{figure*}

We have inspected all the WMAP five-year channels 
(W band at 94 GHz; V band at 61 GHz; Q band at 41 GHz), and performed the same 
analysis described in the main text for each single differencing assembly (DA).
For this study, it is ideal to consider individual DA's rather than their combinations, 
as WMAP beams and noise properties are well-defined within each single DA.
Figure \ref{psigma_map_app} shows the distributions of the rms noise values for all the DA's
considered; note the characteristic bi-modality of these histograms, present in all the channels.
Figure \ref{pDobs_app} shows the distributions of temperatures in the data at different frequencies.
Dotted lines in the figure are Gaussians with the same rms as data, solid 
curves show the predicted distributions when inhomogeneous noise effects are included.
Departures from the Gaussian fits are more significant in the W frequency range.
Figure \ref{number_density_512_app} shows the number density of pixels above (below)    
different temperature thresholds. Dotted lines are theoretical predictions with homogeneous
noise, solid lines include inhomogeneity. Points are measurements from all the WMAP5 DA's.
Finally, Figure \ref{cf_512_nu_0.25_0.75_app} highlights 
some examples of the unweighted correlation functions from the WMAP5
pixel-pixel temperature fields, at $N_{\rm side}=512$. Two
pixel-thresholds are considered ($|\nu|=0.25$ and $|\nu|=0.75$), as
indicated in the panels. Solid curves show the
predictions associated with Gaussian signal plus inhomogeneous
noise. Points are measurements of the clustering of hot and cold
pixels at corresponding temperature thresholds, and shaded areas are the
$1\sigma$ optimal variance errors. 
Essentially, we find that the difference in clustering between hot and cold pixels
is still present in all the other DA's. 
However, our theory is always in good agreement with the clustering of cold pixels,
and in slight disagreement with that of the hot ones, both in the Q and in the V channels.
Therefore the 94 GHz frequency (W band) seems to be the most
discrepant one, with respect to our theoretical predictions.
In particular, we checked our small-scale theory expectations against 
WMAP mock measurements from simulations with identical beam and noise
properties, and found full consistency. Hence, we suggest that 
those small-scale discrepancies may be due to extragalactic dust emission (one
type of point-source contamination), which has a peak at high
frequencies and is not accounted in our analysis.    



\end{document}